\documentclass[10pt,leqno,oneside]{extarticle}
\usepackage{cite}

\usepackage{amsmath, amssymb}
\usepackage{amsfonts}
\usepackage{setspace}
\usepackage{ccfonts} 
\usepackage[varg]{txfonts}
\usepackage{latexsym}
\usepackage{mathrsfs}

\newcommand{\field}[1]{\mathbb{#1}}

\newcommand{\n}{\noindent}
\newcommand{\be}{\begin{equation}}
\newcommand{\ee}{\end{equation}}
\newcommand{\ben}{\begin{displaymath}}

\newcommand{\een}{\end{displaymath}}
\newcommand{\ep}{\hspace{\stretch{1}}$\Box$}

\newcommand{\vs}{\vspace{0.2cm}}

\newcommand{\stt}{{\tilde{\tt s}}}
\newcommand{\st}{{\tt s}}
\newcommand{\hst}{\hat{\st}}
\newcommand{\tbg}{\bar{\tt g}}
\newcommand{\tg}{{\tt g}}
\newcommand{\bvl}{\bigg|}
\newcommand{\sta}{\hat{\st}_{\alpha}}

\newtheorem{Remark}{Remark}
\newtheorem{Definition}{Definition}
\newtheorem{Proposition}{Proposition}
\newtheorem{Theorem}{Theorem}
\newtheorem{Conjecture}{Conjecture}
\newtheorem{Lemma}{Lemma}

\newtheorem{Corollary}{Corollary}
\newtheorem{Example}{Example}

\newtheorem{Notation}{Notation}

\begin{document}

\begin{center}
{\bf\Large Static solutions from the point of view of comparison geometry.}
\end{center}

\begin{center}
{Martin Reiris.}\footnote{e-mail: martin@aei.mpg.de}\\

\vs
\textsc{Max Planck Institute f\"ur Gravitationasphysik. \\ Albert Einstein Institut.\\ Germany.}\\

\vspace{0.7cm}
\begin{minipage}[c]{11cm}
\begin{spacing}{0.8}
\begin{center}
{\bf Abstract.}
\end{center}
{\small We analyze (the harmonic map representation of) static solutions of the Einstein Equations in dimension three from the point of view of comparison geometry. We find simple monotonic quantities capturing sharply the influence of the Lapse function on the focussing of geodesics. This allows, in particular, a sharp estimation of the Laplacian of the distance function to a given (hyper)-surface. We apply the technique to asymptotically flat solutions with regular and connected horizons and, after a detailed analysis of the distance function to the horizon, we recover the Penrose inequality and the uniqueness of the Schwarzschild solution. The proof of this last result does not require proving conformal flatness at any intermediate step.}  
\end{spacing}
\end{minipage}
\end{center}

\begin{center}
\begin{minipage}{11cm}
\tableofcontents
\end{minipage}
\end{center}

\newpage
\section{Introduction.}
In this article we introduce a family of quantities, denoted by ${\mathcal{M}}_{a}$ (where $a$, an arbitrary real number, is the parameter of the family) naturally attached to (integrable) geodesic congruences ${\mathcal{F}}$, of Static Solutions of the Einstein Equations in dimension three. The invariants (which can be seen as a real functions over the range of the congruence) are shown to be monotonic along each of the geodesics of ${\mathcal{F}}$. Moreover whenever ${\mathcal{M}}_{a}$ is stationary along a geodesic $\gamma$ of ${\mathcal{F}}$, then the local geometry along $\gamma$ can be seen to be of Schwarzschild form. In this sense ${\mathcal{M}}_{a}$ measures a certain departure of the given static solution to the Schwarzschild solution. The framework that we will develop out of these invariants
is a natural extension of the standard comparison techniques of Riemannian spaces of non-negative Ricci curvature. However, as we incorporate into ${\mathcal{M}}_{a}$ the influence that the lapse exerts on the Ricci curvature, and, as a result, the monotonicity of ${\mathcal{M}}_{a}$ sharply captures the departure from the Schwarzschild solution (not from the Euclidean space), the framework here developed can be best described as one that compares static solutions to the Schwarzschild solution.  It is thus not peculiar that when the technique is applied to asymptotically flat static solutions with regular and connected horizons, the uniqueness of the Schwazschild solution is achieved with remarkable naturalness. It is worth noting that the novel proof of this central result in General Relativity that we shall provide does not require the intermediate step of proving conformal flatness of previous proofs.  The ideas that we will describe can be interpreted as partial results on the bigger proposal of developing a more complex comparison theory for static solutions in arbitrary dimensions. 

Before continuing with the description of the contents, we briefly introduce static solutions of the Einstein equations and summarize some properties that would place the contents into an adequate perspective.

\subsection{Elements of static solutions.}

A static solution of the Einstein equations in dimension three \footnote{In this article we will restrict to dimension three. Our most important invariant, the quantity ${\mathcal{M}}$ (see later), is monotonic only in dimension three and we do not know, at the moment, a replacement of it to higher dimensions. The static Einstein equations (\ref{SE1})-(\ref{SE2}) are valid in any dimension.}, is given by a triple $(\Sigma,g,N)$ where $\Sigma$ is a smooth Riemannian three manifold possibly with boundary, $g$ is a smooth Riemannian metric and $N$, the {\it Lapse Function}, is a smooth function, strictly positive in ${\rm int (}\Sigma{\rm )}$, and satisfying
\begin{align}\label{SE1}
N Ric&=\nabla\nabla N,\\
\label{SE2}
\Delta N&=0.
\end{align}

\n These equations, note, are invariant under simultaneous but independent scalings on $g$ and $N$. 

The description of static solutions is better separated into local and global properties. From the local point of view, the geometry of static solutions is controlled in $C^{\infty}$ by two weak invariants. This is a direct consequence of Anderson's curvature estimates ~\cite{MR1809792} (applying in dimension three) which are described as follows. Let $(\Omega,g,N)$ be a static solution of the Einstein equations, where $(\Omega,g)$ is a complete Riemannian manifold with or without boundary. Then there is a universal constant $K>0$ such that for any $p\in\Omega$ we have
\be
|Rm|+|\nabla \ln N|^{2}\leq   \frac{K}{dist(p,\partial \Omega)^{2}},
\ee

\n where if $\partial \Omega=\emptyset$ we set $dist(p,\partial \Omega)=\infty$. Note that this shows in particular that the only complete and boundary-less static solution in dimension three is covered (after normalizing $N$ to one) by the trivial solution $(\field{R}^{3},g_{\field{R}^{3}},N=1)$. Anderson's curvature estimates together with the Bishop-Gromov volume comparison and standard elliptic estimates, imply the following {\it interior estimates} for static solutions in dimension three.

\begin{Lemma} {\rm {\bf (Interior's estimates (Anderson)})} Let $\Omega$ be a closed three-dimensional manifold with non-empty boundary $\partial \Omega$. Suppose that $(\Omega,g,N)$ is a static solution of the Einstein equations. Let $p\in \Omega$, let $d=dist(\partial \Omega)$ and let $V_{1}=Vol(B(p,d_{1}))$ for $d_{1}<d$. Then there is $d_{2}(d,d_{1},V_{1})>0$, and for any $i\geq 0$ there are $\Lambda(d,d_{1},V_{1},i)>0$, $I(i,d_{1},V_{1})>0$,  such that $inj(p)\geq I$ and $\| \nabla ^{i}Rm\|_{L^{\infty}_{g}(B(p,d_{2}))}\leq \Lambda$.
\end{Lemma}

\n These interior estimates, in turn imply, as is well known, the control of the $C^{i}_{\{x_{j}\}}$ norm of the entrances $g_{ij}$ of $g$, in suitable harmonic coordinates $\{x_{j}\}$ covering $B(p,d_{2})$, and from them precompactness statements can be obtained.

The global geometry of static solutions instead is greatly influenced by boundary conditions and, in many cases, boundary conditions provide uniqueness. This occurs when, for instance, one assumes that $\partial \Sigma$ consist of a finite set of {\it regular horizons} plus further hypothesis on the asymptotic of $(\Sigma,g)$ at infinity. We will adopt the following definition (see \cite{MR1809792}).

\begin{Definition}
The boundary $\partial \Sigma$ of the smooth manifold $\Sigma$ is a {\it regular horizon} iff $\partial \Sigma$ is a finite union of compact (boundary-less) surfaces $H_{i},i=1,\ldots n$, $\partial \Sigma=\{q/N(q)=0\}$ and at each $H_{i}$ we have $|\nabla N|\bvl_{H_{i}}>0$.
\end{Definition}

\n It follows easily from the static equations (\ref{SE1})-(\ref{SE2}) that every regular horizon $\partial \Sigma$ is totally geodesic and $|\nabla N|$ is constant and different from zero on each component. 

Perhaps the easiest examples of complete solutions with regular horizons are the {\it Flat solutions} that we will denote by the triple $(\Sigma_{F},g_{F},N_{F})$. They have the presentation 
\be
\Sigma_{F}=[0,\infty)\times T^{2},\ N_{F}=r,\ g_{F}=dr^{2}+h_{F},
\ee

\n where $h_{F}$ is a flat metric in $T^{2}$. The family is parameterized by the set of flat metrics in $T^{2}$ (non-isometric). Note that we have demanded that $N$ grows linearly with respect to arc length and with slope one. Of course any $N$ that grows linearly can be scaled to have growth of slope one.  
  
Yet, the prototypical and central examples of static metrics are the Schwarzschild solutions. Recall, the Schwarzschild solution $(\Sigma_{N},g_{S},N_{S})$ of mass $m\geq 0$ has the presentation
\be\label{SchP} 
\Sigma_{S}=[2m,\infty)\times S^{2},\ N_{S}=\sqrt{1-\frac{2m}{r}},\ g_{S}=dr^{2}+r^{2}(1-\frac{2m}{r})d\Omega^{2},
\ee

\n while if $m<0$ the presentation
\be\label{SchP2} 
\Sigma_{S}=(0,\infty)\times S^{2},\ N_{S}=\sqrt{1-\frac{2m}{r}},\ g_{S}=dr^{2}+r^{2}(1-\frac{2m}{r})d\Omega^{2}.
\ee

\n The ``uniqueness of the Schwazschild solution", as in known today and in the form presented below, came as the result of several efforts, starting from the seminal work of Israel in 1967. For the history of the developments which lead to the proof of this important result as well as accurate references we refer to the article \cite{Robinson}. 

\begin{Theorem} {\rm (Schwarzschild's uniqueness ~\cite{Israel}, ~\cite{RobinsonII},~\cite{MR876598})}\label{TRS}
Let $(\Sigma,g,N)$ be a static solution of the Einstein equations of dimension three. Suppose it is asymptotically flat (with one end) and with regular, possibly empty, and possibly disconnected horizon $\partial \Sigma$. Then the solutions is a Schwarzschild solution of non-negative mass. 
\end{Theorem}

Several hypothesis of this theorem can be relaxed still obtaining the same uniqueness outcome. 
For instance suppose there is one end but the hypothesis of asymptotic flatness, or even the topological nature of the end, is withdrawn, then results exist showing that the solution is still one of the Schwazschild family of positive mass. In particular when $N\leq N_{0}<\infty$ but nothing of the end is known, not even the a priori topology, then it can be shown
\footnote{This follows from a combination of results. First observe that $N$ cannot go uniformly to zero over the end, for in such case, as $N$ is harmonic and is zero over the horizon, we would violate the maximum principle. Using the notation in ~\cite{MR1809792} denote by $t(p)$ the $g$-distance from a point $p$ to the horizon $H$. Denote also by $B(H,\bar{t})$ the ball of center $H$ and radius $\bar{t}$, namely $B(H,\bar{t})=\{p/ t(p)<\bar{t}\}$. Now, from Theorem 0.3 (ii) in\cite{MR1809792}, either the end is asymptotically flat or small in the sense that $\int^{\infty}\frac{1}{A(\partial B(H,\bar{t}))}d\bar{t}=\infty$. Assume $N\leq N_{0}$. Consider $f=N_{0}+1-N$. Then $\Delta f=0$ and $\Delta \ln f=-|\nabla \ln f|^{2}$. Define $F(t):=\int_{B(H,t)\setminus B(H,\bar{t}_{1})}|\nabla \ln f|^{2}dV$. For $t_{1}$ small, we have $\int_{\partial B(H,t_{1})}g(\nabla \ln f,n_{in})dA>0$, where $n_{in}$ is the unit normal to $\partial B(H,t_{1})$ pointing inwards to the ball. Using this fact, integrating $\Delta \ln f=-|\nabla \ln f|^{2}$ over $B(H,t)\setminus B(H,\bar{t}_{1})$ and using Cauchy-Schwarz one easily deduce the inequality $F'/F^{2}\geq 1/A$. From it one gets 
$1/F(t)\leq 1/F(t_{2})-\int_{t_{2}}^{t}\frac{1}{A}d\bar{t}$, where $t_{2}>t_{1}$. Thus if the end is small, one would get $F=\infty$ at a finite distance form $H$, which is not possible.} that the solution is indeed a Schwarzschild solution. The same occurs when it is known that outside a compact set, each end is homeomorphic to $\field{R}^{3}$ minus a ball\footnote{arXiv:1002.1172} and over there the metric $N^{2}g$ is complete, which occurs for example when $N\geq N_{0}>0$. In all these generalizations, which are important for deeper understanding of Einstein's theory, it is assumed that the space $(\Sigma,g)$, as a metric space, is complete. 

We feel that the following broader conjecture may be accessible.
\begin{Conjecture}  Let $(\Sigma,g,N)$ be a complete solution of the Static Einstein equations with regular but possibly disconnected (non-empty) horizon $\partial \Sigma$. Suppose that the conformal metrics $N^{2}g$ and $N^{-2}g$ are complete outside given domains of compact closure on each end of $(\Sigma,g)$. Then the solution is either a Schwarschild solution or a flat solution.
\end{Conjecture} 

\n Observe that no assumption is made on the topology of the ends.
   
When boundary data is prescribed, and is not the data of a regular horizon, and the hypothesis of asymptotic flatness is kept, then much less is known about the existence of solutions although a conjecture \cite{MR1957036} and partial results do exist\footnote{arXiv:0909.4550} under some hypothesis.  In whatever case, Dirichlet-type of problems for the Einstein equations are interesting from physical and mathematical reasons. A theory, a highly necessary task, is still lacking.  

The Schwarzschild family is unique, but, why?. Are the present proofs satisfactory as an answer to this question?. Do we need to place the problem of the uniqueness of the Schwarzschild family into a larger one to understand it better?. Which one would be that bigger perspective?. Could it be a Dirichlet-type of theory for the Static Einstein equations?. Despite all the accumulated knowledge, some aspects of the uniqueness of the Schwarzschild solutions remains (to us) somehow mysterious. The present work would try to clarify the phenomenon from the perspective of comparison geometry. It is worth finally to remark that there are yet further reasons of why it is important to have different proofs and points of view regarding Theorem \ref{TRS}. Just mention that the elusive and yet inconclusive notions of localized energy or the even more conjectural notion of entropy may have to do and could be better clarified with different understandings of the Schwazschild uniqueness.
    
\subsection{${\mathcal{M}}_{a}$ and comparison geometry.}    

\vs
The idea underlying the technique that we will describe is rather simple. First, and most important, we will work in the {\it harmonic map representation} of static solutions. Namely, instead of working with the variables $(g,N)$ we will work with the variables $(\tg,{\tt N})=(N^{2}g,\ln N)$. The Einstein equations (\ref{SE1})-(\ref{SE2}) now become 
\begin{align}
{\tt Ric}&=2d {\tt N}\otimes d{\tt N},\\ 
\Delta {\tt N}&=0.
\end{align}

\n It is apparent from here that ${\tt Ric}\geq 0$, which is a quite central property. Consider now a congruence of geodesics (or geodesic segments) ${\mathcal{F}}$ for the metric $\tg$ minimizing the distance from any of their points to a (hyper)-surface ${\mathcal{S}}$. Thus any geodesic in ${\mathcal{F}}$ has an initial point in ${\mathcal{S}}$. We will assume the geodesics (or geodesic segments) are inextensible beyond their last point or that the last point is the point on $\gamma$ where $\gamma$ stops to be length minimizing to ${\mathcal{S}}$. It can be that such last point does not exists in which case the geodesic ``ends" at ``infinity".   It is known that the {\it Cut locus} ${\mathcal{C}}$, namely the set of last points of the geodesics in the congruence is a closed set of measure zero. Outside ${\mathcal{C}}$ the distance function to ${\mathcal{S}}$ is a smooth function with gradient of norm one. Given a point $p$ in $\Sigma$, we will denote by $\st(p)$ the distance from $p$ to ${\mathcal{S}}$. Consider now a point $p$, not in ${\mathcal{C}}$ and not in ${\mathcal{S}}$ and around it consider the smooth surface formed by the set of points which have the same distance to ${\mathcal{S}}$ than $p$ (the equidistant surface or the level set of the distance function). The second fundamental form of such surface in the outgoing direction (from ${\mathcal{S}}$) at $p$ will be denoted by $\Theta(p)$ or simply $\Theta$. The mean curvature will be $\theta(p)=tr_{h(p)}\Theta(p)$ where $tr_{h(p)}\Theta(p)$ means the trace of $\Theta(p)$ with respect to the induced two-metric in the surface or level set. Thus we can think $\theta$ as a function along geodesics $\gamma$ in ${\mathcal{F}}$. The mean curvature satisfies the important {\it focussing equation} or {\it Riccati equation} along the geodesics $\gamma$        
\be\label{FOCEQ}
\theta '=-|\Theta|^{2}-{\tt Ric}(\gamma',\gamma')=-\frac{\theta^{2}}{2}-{\tt Ric}(\gamma',\gamma')-|\hat{\Theta}|^{2}.
\ee

\n Above, $'$ denotes derivative with respect to arc length and $\hat{\Theta}$ is the traceless part of $\Theta$. Recall that
\ben
\Delta \st=\theta.
\een

\n Thus any estimate on $\theta$ obtained out of the focussing equation serves as an estimate on the Laplacian of the distance function. 

For instance if ${\tt Ric}\geq 0$ then standard estimates in comparison theory follow by discarding the last two terms in equation (\ref{FOCEQ}) and integrating the inequality $\theta'\leq -\theta^{2}/2$. If $\st$ is the distance function to a point, or, the same, the distance function to the boundary of a small geodesic ball plus the radius of the ball, one gets (Calabi 1958), $\theta\leq 2/\st$ and 
\ben
 \Delta \st\leq 2/\st,
\een 
 
\n  everywhere and in the barer sense (~\cite{MR2243772}, pg. 262). Comparison estimates on areas and volumes of geodesic balls are obtained from 
\ben
\frac{dA'}{dA}=\theta,\ \ dV'=dA.
\een

\n where $dA$ is the element of area of the equidistant surfaces to ${\mathcal{S}}$ and $dV$ is the element  of volume enclosed by $dA$. 

The situation we face is similar in that the Ricci curvature is non-negative, but this time the structure of the Ricci curvature is explicitly given. By incorporating ${\tt Ric}$ as part of the focussing inequality, namely considering
\ben
\theta'\leq -\frac{\theta^{2}}{2}-2{\tt N}'^{2},
\een

\n we will obtain a sharp estimate for $\theta$. We will show that for any real number $a$ the quantity 
\ben
{\mathcal{M}}_{a}=(\frac{\theta}{2}(\st+a)^{2}-(\st+a))N^{2},
\een

\n is monotonically decreasing (Proposition \ref{MMP}) along any geodesic of the congruence and is stationary if and only if the geometry along the geodesic is of Schwarzschild form (Proposition \ref{COM}). Thus we get the estimate
\ben
\theta\leq \frac{2}{\st+a}(1+\frac{{\mathcal{M}}_{0}}{(\st+a)^{2}N^{2}}).
\een      

\n where ${\mathcal{M}}_{0}$ is the value of ${\mathcal{M}}_{a}$ at the start, on ${\mathcal{S}}$, of the geodesic. The fundamental set of equations out of which comparison estimates can be obtained is therefore
\begin{align}
&\theta\leq \frac{2}{\st+a}(1+\frac{{\mathcal{M}}_{0}}{(\st+a)^{2}N^{2}}),\label{FE1}\\
&\Delta\st=\theta,\ \frac{dA'}{dA}=\theta,\ dV'=dA,\label{FE2}\\
&\Delta \ln N=0\label{FE3}.
\end{align}

\n To use these set of equations efficiently one must first use the system
\begin{align*}
&\Delta\st\leq \frac{2}{\st+a}(1+\frac{{\mathcal{M}}_{0}}{(\st+a)^{2}N^{2}}),\\
&\Delta \ln N=0,
\end{align*}

\n together with additional boundary data on $N$ and ${\mathcal{M}}_{0}$. For the case of the application to the uniqueness of the Schwarszchild solutions, that we carry out later, the substantial information that is extracted out of this system is, in a sense, concentrated in Theorem \ref{DC}, where a distance comparison result is established between $\st$ and $\hst=2mN^{2}/(1-N^{2})$. 

From the point of view of areas and volumes comparisons, we note that, by using equations (\ref{FE1})-(\ref{FE3}), the expression 
\ben
\frac{dA}{dA_{0}}exp(\int_{\st_{0}}^{\st}\frac{2}{\bar{\st}+a}(1+\frac{{\mathcal{M}}_{0}}{\bar{\st}+a)^{2}N^{2}})d\bar{\st}),
\een

\n is seen to be monotonically decreasing too. From it and $dV'=dA$ suitable information on the growth of areas and volumes of geodesic balls (with center ${\mathcal{S}}$) can be obtained. These type of estimates will play an important role in the proof of the uniqueness of the Schwarzschild solutions in Section \ref{AVC}. 

Yet, the structure of the harmonic-map representation of the Einstein equations is richer than the information contained in the system (\ref{FE1})-(\ref{FE3}). Indeed, Weitzenb\"och's formula for the static equations 
\ben
\frac{1}{2}\Delta |\nabla f|^{2}=|\nabla\nabla f|^{2}+<\nabla \Delta f,\nabla f>+2<\frac{\nabla N}{N},\nabla f>^{2},
\een

\n valid for any function $f$, together with equation (\ref{FE3}) can provide useful estimates on functions of the form $f=f(N)$. They, in turn, provide useful information on $N$. These estimates, is worth remarking, have nothing to do with the distance function. The most obvious consequence of Weitzenb\"ock's formula comes out when we chose $f=\ln N$. In this case we obtain
\ben
\frac{1}{2}\Delta |\nabla \ln N|^{2}=|\nabla\nabla \ln N|^{2}+2|\nabla \ln N|^{2}.
\een

\n In applications to the uniqueness of the Scharzschild solutions, we will use however the Weintzenb\"ock formula with the choice $f=\hst=2mN^{2}/(1-N^{2})$. This will provide the important estimate $|\nabla \hst|\leq 1$ in Section \ref{FPCDL}, which, as we will see, it is necessary to close up the proof of the uniqueness of the Schwarzschild solutions. 

It is worth remarking at this point that many of the techniques here developed carry over the much bigger family of metrics and potentials satisfying 
\begin{align*}
&{\tt Ric}\geq 2d{\tt N}\otimes d{\tt N},\\
&\Delta {\tt N}\geq 0,\ \ 0<N<1.
\end{align*}

To show the applicability of equations (\ref{FE1})-(\ref{FE2}), as we said before, we will fully analyze from this perspective asymptotically flat static solutions with regular and connected horizons and recover Theorem \ref{TRS}. It is worth remarking the naturalness from which the uniqueness of the Schwarzschild solutions will come out of these comparison techniques. Despite of that, the required analysis will be somewhat extensive. 
To prove Theorem \ref{TRS} we carefully compare the distance function to the horizon, $\st$, to the function $\hst$, through the set of equations (\ref{FE1})-(\ref{FE2}). The final goal to achieve is to show the equality $\st=\hat{\st}=2mN^{2}/(1-N^{2})$ from which it will follow that ${\mathcal{M}}_{2m}$ has to be stationary along any length-minimizing geodesic to $H$ (in this case the integral lines of $\nabla \hst$) and equal to $m$. It then follows, from the sharpness of the monotonicity of ${\mathcal{M}}$ that the solution has to be a Schwazschild solution (of positive mass). Note a technical aspect however. As $N=0$ over the horizon $H$, the metric $\tg=N^{2}g$ is singular there. Although this will make the analysis technically delicate, a satisfactory remedy is found if one replaces $H$ by a sequence $H_{\Gamma_{i}}=\{N=\Gamma_{i}\}$ of the $\Gamma_{i}$-level set of $N$ ($\Gamma_{i}\downarrow 0$), approaching $H$, and perform then a limit analysis. This circumvention of the singularity at the horizon will appear often in the reasonings.   

It is worth noting that, at the moment, we do not know how to obtain Theorem \ref{TRS}, when the horizon is not connected. The exact reproduction of the arguments that leads to the proof of Theorem \ref{TRS} for connected horizons, applied to the case of non-connected horizons, give interesting results, which are not difficult to obtain but that will not be given here.

We will now give guidelines of the structure of the article. In Section \ref{CON2} we introduce and discuss the main monotonic quantity ${\mathcal{M}}$, give explicit examples of the monotonicity and discuss the stationary case. This section is the core of the article. The other sections discuss further properties of ${\mathcal{M}}$ and applications. In particular in Section \ref{CON3} we start the discussion of asymptotically flat solutions with regular and connected horizons. In Section \ref{SMRH} we study ${\mathcal{M}}$ over regular horizons. In Section \ref{SAF} we recall the notion of asymptotic flatness and cite a classical result \cite{MR608121} on the possibility to chose special coordinates at infinity in static solutions displaying precisely the Scwarzschild-type of fall off. The existence of such coordinate system $\{\bar{x}\}$ will be central. In Section \ref{SCDL} we introduce the important notion of {\it coordinate-distance} lag, measuring a mismatch between the distance from a point $p$ to the horizon, $\st(p)$, and the  coordinate distance $|\bar{x}(p)|$. In Section \ref{SDC} we discuss our first substantial result. We prove a {\it distance comparison} result (Theorem \ref{DC}) between $\st$ and $\hat{\st}=2m N^{2}/(1-N^{2})$. To achieve it, we must show first that the inequality 
\ben
\Delta\st\leq \frac{2}{\st+a}(1+\frac{{\mathcal{M}}_{0}}{(\st+a)^{2}N^{2}}),
\een

\n holds in a barer sense all over the manifold $\Sigma$. This is done in Proposition \ref{LSH}. Without that tool, the comparison result would not be possible to achieve. Using that we show in Section \ref{SPI} that the Penrose inequality $A\leq 16\pi m^{2}$, where $A$ is the area of the connected horizon and $m$ is the ADM mass, must hold. In Section \ref{OSPI} we show, again using the distance comparison,  that the opposite Penrose inequality must hold, namely $A\geq 16\pi m^{2}$. Thus, after Section \ref{OSPI} we would have proved that $A=16\pi m^{2}$. Despite the strong implications of this inequality, the uniqueness of the Schwarzschid solutions requires further analysis. This is carried out in Section \ref{USS}. Indeed it is in this section that it is proved that $\st=\hat{\st}$. This follows from a further study of the coordinate-distance lag in Section \ref{FPCDL} where it is shown that it must be zero. An elaboration of the area and volume comparison in Section \ref{AVC} finishes the analysis of all the elements of the proof which is summarized in Theorem \ref{TUSFIN}.  Further explanations on the contents and strategies are given at the beginning of each Section. 

We will use alternatively the notation $(\Sigma,g,N)$ or $(\Sigma,\tg,{\tt N})=(\Sigma,\tg,\ln N)$, used according to which representation is best suited to describe a claim or a statement. When we say that $(\Sigma, \tg, \ln N)$ is an asymptotically flat static solution with regular and connected horizon, we mean that $(\Sigma,g,N)$ is an asymptotically flat static solution with regular horizon as was described before, but that we will working in its harmonic map representation.   

\section{A comparison approach to static solutions in the harmonic map representation.}\label{CON2}

Let $(\Sigma,{\tt g},\ln N)$ be a static solution in the harmonic representation. To every oriented integrable congruence ${\mathcal{F}}$ of ${\tt g}$-geodesics, we will associate a family of real functions $\{{\mathcal{M}}_{a},a\in \field{R}\}$ defined over the range of ${\mathcal{F}}$. We will show that, fixed $a$, ${\mathcal{M}}_{a}(\gamma ({\tt s}))$ is monotonically decreasing for any $\gamma\in {\mathcal{F}}$ (${\tt s}$ is the ${\tt g}$-arc length, increasing in the positive direction). This central fact will follow by making use of the focusing equation (\ref{FOCEQ}). The definition of ${\mathcal{M}}_{a}$ and the proof of its monotonicity are given in the Proposition below. To avoid excessive notation we will use the following convention in the notation: for every function $f$ defined over the range of ${\mathcal{F}}$ (for example $f=\theta$ or $f=N$) and $\gamma\in {\mathcal{F}}$ we will write $f:=f(\gamma(s))$ and $df(\gamma(s))/ds:=df/ds:=f'$. Also, for the same reason of economy and simplicity, we will suppress the sub-index $a$ and write simply ${\mathcal{M}}$.  

\begin{Proposition}\label{MMP}
Let ${\mathcal{F}}$ be an oriented integrable congruence of geodesics. Let $\gamma({\tt s})$, ${\tt s}\in [{\tt s}_{0},{\tt s}_{1}]$ be a geodesic in ${\mathcal{F}}$. Let $a$ be a real number and let $\stt=a+\st$. Then we have
\be\label{MME}
((\frac{\theta}{2}\stt^{2}-\stt)N^{2})'=-\stt^{2}N^{2}\frac{|\hat{\Theta}|^{2}}{2}-(\stt\frac{\theta}{2}-1 -\stt\frac{N'}{N})^{2}N^{2}.
\ee

\n Therefore, fixed any real number $a$, the quantity ${\mathcal{M}}=(\frac{\theta}{2}\stt^{2}-\stt)N^{2}$ is monotonically decreasing along any $\gamma\in {\mathcal{F}}$ (the notation ${\mathcal{M}}$ accounts for ``{\it mass}''). 
\end{Proposition}
 
\n {\it Proof:}

\vs
We compute
\ben
((\frac{\theta}{2}\tilde{\st}^{2}-\tilde{\st})N^{2})'=\frac{\theta'}{2}\tilde{\st}^{2}N^{2}+\theta \tilde{\st}N^{2}-N^{2}+\theta\tilde{\st}^{2}N N'-\tilde{\st}2N N'.
\een   
 
\n We use now the focusing equation (\ref{FOCEQ}) to get
\ben
((\frac{\theta}{2}\tilde{\st}^{2}-\tilde{\st})N^{2})'=-\frac{|\hat{\Theta}|^{2}}{2}\tilde{\st}^{2} N^{2} - \frac{\theta^{2}}{4}\tilde{\st}^{2} N^{2} -\tilde{\st} N'^{2} + \theta \tilde{\st}N^{2}-N^{2}+\theta\tilde{\st}^{2}N N'-\tilde{\st}2N N'.
\een    
 
\n The six terms following the first on the right hand side of this expression can be arranged as $-(\tilde{\st}\theta/2-1-N'/N)^{2}N^{2}$, thus obtaining (\ref{MME}).\ep
 
\begin{Example} {\rm ({\it The Schwarzschild case.})\label{EXA1} Consider a Schwarzschild metric of mass $m$, of arbitrary sign, in the presentations, according to the sign of the mass, of equations (\ref{SchP}) or (\ref{SchP2}). Note that ${\tt g}=dr^{2}+r^{2}(1-2m/r)d\Omega^{2}$ and $N^{2}=(1-2m/r)$. For any given point $q$ in $S^{2}$ consider the ray $[2m,\infty)\times \{q\}$ (if $m\geq 0$) or $(o,\infty)\times \{q\}$ (if $m<0$) parameterized by the arc length $\st=r-2m$ of $\st=r$ (respectively to the sign of the mass). In either case we compute the mean curvature $\theta$ as
\be
\theta=\frac{2}{r}+\frac{2m}{r(r-2m)}=\frac{2}{r}\frac{(r-m)}{r-2m}.
\ee

\n Let $b=a-2m$ if $m\geq 0$ and $b=a$ if $m<0$, then the quantity ${\mathcal{M}}$ has the following form,
\ben
{\mathcal{M}}=(\frac{1}{r}\frac{r-m}{r-2m}(r+b)^{2}-(r+b))(1-\frac{2m}{r})=((m+b)-\frac{m b}{r})(1+\frac{b}{r}).
\een

\n independently of the sign of the mass. Taking the derivative with respect to arc length and rearranging terms we obtain 
\ben
{\mathcal{M}}'=\frac{b^{2}}{r}(\frac{2m}{r}-1),
\een

\n which is explicitly non-positive independently of the sign of the mass. This shows the monotonicity of ${\mathcal{M}}$ for any value of $b$. Note that $lim_{\st \rightarrow \infty}{\mathcal{M}}=m+b$. Observe too that when $b=0$, i.e. $a=2m$, then ${\mathcal{M}}$ is constant and equal to $m$.}\ep 
\end{Example}

\begin{Example} {\rm ({\it The flat solutions.}) For a flat solution $(\Sigma_{F},g_{F},N_{F})$ we have ${\tt g}=r^{2}dr^{2} +r^{2}h_{F}$. Making $r^{2}/2=s$ we get ${\tt g}=ds^{2}+2sh_{F}$ and $N^{2}=2s$ where $s>0$. For any point $p$ in $T^{2}$ consider the ray $[0,\infty)\times {p}$. Consider the congruence of geodesics conformed by all these rays. The mean curvature is calculated as $\theta=1/s$. Thus for any real number $a$ we have 
\be
{\mathcal{M}}=(\frac{1}{2s}(s+a)^{2}-(s+a))2s=-s^{2}+a^{2},
\ee

\n which is monotonically decreasing in the domain of $s$, namely $(0,\infty)$. 

Note that for the ``dual'' solution $(\Sigma_{F},g_{F},1/N_{F})$ we have, for any real number $a$, the expression ${\mathcal{M}}=-1/4 +a^{2}/s^{2}$ which is monotonically decreasing in the domain of $s$, namely $(0,\infty)$. Note that when $a=0$ then ${\mathcal{M}}$ is stationary and equal to $-1/4$.}\ep
\end{Example}

\vs
The next proposition discusses the case when ${\mathcal{M}}_{a}$ is stationary.

\begin{Proposition}\label{COM}
Let ${\mathcal{F}}$ be an oriented and integrable congruence of geodesics. When, for a given $a$, ${\mathcal{M}}$ is constant along a geodesic segment $\gamma(\st)$, $\st\in [\st_{1},\st_{2}]$ then along $\gamma$ we have
\be\label{SM1}
\hat{\Theta}=0,
\ee
\n and
\be\label{SM2}
N^{2}=N_{0}^{2}+2\frac{{\mathcal{M}}_{0}}{\st_{0}+a}-2\frac{{\mathcal{M}}_{0}}{\st+a},
\ee

\n where $N_{0}$ and ${\mathcal{M}}_{0}$ are the values of $N$ and ${\mathcal{M}}$ at $\st=\st_{0}\in (\st_{1},\st_{2})$. We also obtain
\be\label{SM3}
\theta=\frac{2}{\st+a}+2\frac{{\mathcal{M}}_{0}}{(\st+a)^{2}N^{2}}.
\ee   
\end{Proposition}

\n {\it Proof:} 

\vs
If along a geodesic $\gamma$ the value of ${\mathcal{M}}$ remains constant, then the right hand side of (\ref{MME}) must be identically zero. This implies that 
\ben
\hat{\Theta}=0,
\een
\n which shows (\ref{SM1}), and also implies that 
\ben
\tilde{\st}\frac{\theta}{2}-1-\tilde{\st}\frac{N'}{N}=0.
\een

\n Multiply now this expression by $\tilde{\st}$ and rearrange it as
\be\label{TEM}
\tilde{\st}^{2}\frac{\theta}{2}-\tilde{\st}=\tilde{\st}^{2}\frac{N'}{N}.
\ee

\n Recall that ${\mathcal{M}}=(\tilde{\st}^{2}\frac{\theta}{2}-\tilde{\st})N^{2}$. Using this expression, the equation (\ref{TEM}), and (because we are assuming that ${\mathcal{M}}$ is constant) writing ${\mathcal{M}}={\mathcal{M}}_{0}={\mathcal{M}}(\st_{0})$, we obtain 
\ben
{\mathcal{M}}_{0} =\tilde{\st}^{2}N N'=\tilde{\st}^{2} \frac{(N^{2})'}{2}.
\een

\n Moving $\tilde{\st}^{2}$ to the denominator of the left hand side and integrating (in $\st$) from $\st=\st_{0}$ to $\st$ we obtain (\ref{SM2}). To obtain (\ref{SM3}) solve for $\theta$ in $(\tilde{\st}^{2}\frac{\theta}{2}-\tilde{\st})N^{2}={\mathcal{M}}_{0}$.
\ep 

\begin{Remark} {\rm (Further remarks to Proposition \ref{COM})
Observe form Proposition (\ref{COM}) that (if for some number $a$) ${\mathcal{M}}$ is constant along a geodesic $\gamma$ of infinite length and $lim_{\st\rightarrow \infty} N(\gamma(\st))=1$, then making the change of variables $r=\st+a$ in (\ref{SM2}) and (\ref{SM3}) we obtain, along $\gamma$, the expressions
\ben
N^{2}(r)=1-\frac{2{\mathcal{M}}_{0}}{r},
\een
\ben
\theta=\frac{2}{r}+\frac{2{\mathcal{M}}_{0}}{r(r-2{\mathcal{M}}_{0})}=\frac{2}{r}\frac{(r-{\mathcal{M}}_{0})}{r-2{\mathcal{M}}_{0}}.
\een

\n and, including (\ref{SM1})
\ben
\hat{\Theta}=0,
\een

\n which, comparing with Example \ref{EXA1}, are exactly of Schwarzschild form if we identify ${\mathcal{M}}_{0}$ with ``a" ADM mass $m$. Moreover if $\gamma$ is defined on $(\st_{0}=0,\infty)$ and $\gamma(\st_{0})$ ``lies'' on a ``horizon'' ($\lim_{\st\rightarrow 0} N(\gamma(\st))=0$), then $N_{0}=0$ and $1=N_{0}^{2}+2{\mathcal{M}}_{0}/(\st_{0}+a)=2m/a$. Therefore $a=2m$ and $\st=r-2m$. Note that $m={\mathcal{M}}_{0}$ cannot be negative otherwise $\theta$ reaches infinity for an $\st\in (0,\infty)$ (thus the whole $\gamma$ cannot belong to ${\mathcal{F}}$). Thus we establish the same relation $\st=r-2m$ as in a Schwarzschild solution of positive mass. 
On the other hand if $\gamma$ is defined on $(\st_{0}=0,\infty)$ and $\gamma(\st_{0})$ ``lies'' on a ``naked singularity'' ($\lim_{\st\rightarrow 0}N(\gamma(\st))=+\infty$), then $a=0$ and $\st=r$. Note that in this case $m={\mathcal{M}}_{0}$ must be negative otherwise $N^{2}=1-2m/r=1-2m/\st$ gets negative for small $\st>0$. Thus we establish the same relation $\st=r$ as in a Schwarzschild solution of negative mass.}
\end{Remark}

\begin{Remark}
There are several ways to include the summand $-2(\dot{N}/N)^{2}$ to obtain an estimation on the growth of $\theta$. The following Proposition, whose proof is left to the reader, is one such instance. Although we will not use it for the rest of the article, it illustrates very well, the many ways in which the focussing equation can be used to extract geometric information. 

\begin{Proposition} Let $\theta$ be the mean curvature of the integrable congruence ${\mathcal{F}}$. Let $\gamma({\tt s})$, ${\tt s}\in [{\tt s}_{0},{\tt s}_{1}]$ be in ${\mathcal{F}}$. Then we have

\begin{enumerate}
\item $\theta N^{2}$ is monotonically decreasing, namely $(\theta N^{2})\dot{}\leq -(\frac{\theta N}{\sqrt{2}}-\sqrt{2}\dot{N})^{2}$. Therefore we have $\theta\leq \theta_{0}(N_{0}/N)^{2}$, where $\theta_{0}=\theta({\tt s}_{0})$ and $N_{0}=N({\tt s}_{0})$. 

\item Suppose that $\theta({\tt s})>0$ for all ${\tt s}$ in $[{\tt s}_{0},{\tt s}_{1}]$. Then we have
\be\label{FE}
\theta({\tt s})\leq \frac{1}{\frac{1}{\theta_{0}}+\frac{{\tt s}-{\tt s}_{0}}{2}+\frac{1}{2\theta_{0}^{2}N_{0}^{4}}\frac{(N^{2}-N_{0}^{2})^{2}}{({\tt s}-{\tt s}_{0})}}.
\ee

\n As $\theta$ is monotonically decreasing the same formula holds for all ${\tt s}$ in the domain where ${\tt \gamma}$ is length minimizing provided only $\theta_{0}>0$. 

\end{enumerate}\ep
\end{Proposition}

Equation (\ref{FE}) clearly displays the influence of the Lapse $N$ in the focussing of geodesics beyond the natural focussing that comes out of the non-negativity of the Ricci curvature. Equation (\ref{FE}) can serve, in particular, to obtain information on the relationship between volume growth of tubular neighborhoods of a horizon and the growth of $N$ from it. 
\end{Remark}

\section{Applications to asymptotically flat static solutions with regular and connected horizons.} \label{CON3}

In this section we show that any asymptotically flat static solution with regular and connected horizon must satisfy the Penrose inequality. This is proved in Section \ref{SPI}. Separately, in Section \ref{OSPI} we will prove that one such solution must satisfy the opposite Penrose inequality and that the horizon must be geometrically round. This will lead us into the verge of proving Theorem \ref{TRS} which is carried out in Section \ref{USS}. To achieve the inequalities some preliminary material is introduced in Sections \ref{SMRH}, \ref{SAF}, \ref{SCDL} and \ref{SDC}. In Section \ref{SMRH} we compute ``the value of ${\mathcal{M}}$'' for the ``congruence of geodesics emanating perpendicularly to $H$'' (note that $\tg$ is singular on $H$) which we will be usied crucially in the other Sections. Technically we will elude the fact that $\tg$ is singular on $H$ by considering instead of $H$ suitable sequences $\{H_{\Gamma_{i}}\}$ of two-surfaces approaching $H$ as $i\rightarrow \infty$. In this way ``the value of ${\mathcal{M}}$ over $H$'' will be defined as a limit. Similarly we will define $\st(p):=dist_{\tg}(p,H):=\lim_{i\rightarrow \infty} dist_{\tg}(p,H_{\Gamma_{i}})$.  In Section \ref{SMRH} we recall the notion of {\it Asymptotic Flatness} and introduce, following \cite{MR608121}, a coordinate system adapted to asymptotically flat static solutions that will be very useful later. In Section \ref{SCDL} we introduce the notion of {\it Coordinate-Distance Lag} which is necessary to prove, in Theorem \ref{DC}  of Section \ref{SDC},  a central {\it Distance Comparison}  where we establish a lower bound for the $\tg$-distance function to the horizon (the function $\st$) in terms of a certain function of $N$, $m$ and $A$ (the function $\hat{\hat{\st}}$). For any divergent sequence of points $\{p_{i}\}$ the coordinate-distance lag associated to $\{p_{i}\}$ is defined as $\bar{\delta}(\{p_{i}\})=\limsup \st(p_{i})-r(p_{i})+2m$, where $r=|\bar{x}|$ and $\{\bar{x}=(x_{1},x_{2},x_{3})\}$ is the coordinate system introduced in Section \ref{SAF} and it will be seen to be $\bar{\delta}(\{p_{i}\})=\limsup \st(p_{i})-\hst(p_{i})$. The Penrose inequality in Section \ref{SPI} is then proved by  showing first, using a standard comparison of mean curvatures, that if $P:=A/(16\pi m)>1$ (i.e. the Penrose inequality does not hold) then there is a divergent sequence whose coordinate-distance lag is non-negative  (Corollary \ref{CDC}) and on the other hand proving, using the distance comparison of Section \ref{SDC}, that if $P>1$ then the coordinate-distance lag must be negative for any divergent sequence (Proposition \ref{PDC}). This reaches a contradiction. To prove the opposite Penrose inequality it is shown that the Gaussian curvature $\kappa$ of $H$ must satisfy $\kappa\geq 4(4\pi m/A)^{2}$ to prevent a violation of the distance comparison near the horizon. integrating this inequality over $H$ and using Gauss-Bonnet the opposite Penrose inequality is achieved. As a byproduct of both inequalities one obtains that the horizon must be geometrically round, namely that $\kappa=4\pi/A$.  

\subsection{The value of ${\mathcal{M}}$ over regular horizons.}\label{SMRH}

Let $(\Sigma,\tg,\ln N)$ be an static solution and let $H$ be a regular and connected horizon. Consider an embedded (orientable) surface ${\mathcal{S}}\subset \Sigma\setminus H$. Let $n_{1}$ and $n_{2}$ be the two unit-normal vector fields to ${\mathcal{S}}$. As we noted before if ${\mathcal{F}}$ is the congruence of geodesics emanating perpendicularly to ${\mathcal{S}}$ and following one of the perpendicular directions to ${\mathcal{S}}$, say $n_{1}$, then the mean curvature $\theta$ of the congruence ${\mathcal{F}}$ over ${\mathcal{S}}$ is equal to the mean curvature of the surface ${\mathcal{S}}$ in the direction of $n_{1}$. Now to define ${\mathcal{M}}$ over $H$ (where $\tg$ is singular) for the ``congruence of geodesics emanating perpendicular to $H$'' we will calculate ${\mathcal{M}}$ over a suitable sequence of surfaces and then take the limit as the surfaces approache $H$. Such calculation is performed in the paragraphs below. The following Notation will be used in this Section and those that follow.

\begin{Notation}\label{NOT1}
Let $\Gamma_{0}$ be a number sufficiently small in such a way that for any $\Gamma\leq \Gamma_{0}$, $\Gamma$ is a regular value for the lapse $N$ and the set $H_{\Gamma}:=\{N=\Gamma\}$ is isotopic to $H$ (note that $|\nabla N|\neq 0$ over a regular horizon $H$). One such $\Gamma_{0}$ will be called regular. For any two $\Gamma<\bar{\Gamma}$ denote by $\Omega_{\Gamma,\bar{\Gamma}}$ the closed region enclosed by $H_{\Gamma}$ and $H_{\bar{\Gamma}}$. The region enclosed by $H_{\Gamma}$ and $H$ will be denoted by $\Omega_{H,H_{\Gamma}}$. 
\end{Notation}

\n Let $\{\Gamma_{i}\}_{i=1}^{i=\infty}$ be a sequence such that $\Gamma_{i}\downarrow 0$ and $\Gamma_{i}\leq \Gamma_{0}$ with $\Gamma_{0}$ as in Notation \ref{NOT1}. Define 
\ben
{\mathcal{M}}_{H}:= \lim_{\Gamma_{i}\rightarrow 0} (\frac{\theta}{2}a^{2}-a)N^{2}|_{H_{\Gamma_{i}}}.     
\een

\n The next Proposition shows the limit above exists (so it is well defined) and is always constant over $H$. Define $|\nabla N|_{H}=|\nabla N|_{g}|_{H}$.

\begin{Proposition}\label{VMH}
Let $(\Sigma,g,N)$ be a static solution with regular horizon $\partial \Sigma$. Let $H$ be a connected component of $\partial \Sigma$. Then we have 
\be\label{MH}
{\mathcal{M}}_{H}=|\nabla N|_{H}a^{2}. 
\ee

\end{Proposition}

\n {\it Proof:} 

\vs
Denote (as we have done before) by $\theta$ the mean curvature of $H_{\Gamma}$ with respect to ${\tt g}$ and $\theta_{g}$ the mean curvature with respect to $g$. From the conformal relation ${\tt g}=N^{2}g$ we know that
\ben	
\theta=\frac{\theta_{g}}{N}+2\frac{n(N)}{N^{2}},
\een

\n where $n(N)$ is the normal derivative of $N$ in the outgoing direction (outgoing to $\partial \Omega_{H,H_{\Gamma_{i}}}$ and $n$ a unit vector with respect to $g$). Thus we get
\ben
(\frac{\theta}{2}a^{2}-a)N^{2}=a^{2}n(N)+\frac{a^{2}\theta_{g}N}{2}-aN^{2}.
\een

\n We get equation $(\ref{MH})$ in the limit when $\Gamma_{i}\rightarrow 0$. \ep

\subsection{Asymptotically flat static solutions.}\label{SAF}

We will use a useful characterization of asymptotically flat static solutions $(\Sigma, \tg, \ln N)$ due to Beig and Simon \cite{MR608121}. Following \cite{MR608121} we say that $(\Sigma,\tg,\ln N)$ is {\it asymptotically flat} iff there is a coordinate system $\{{\bar{x}=(x_{1},x_{2},x_{3})}$\ {\rm with}\ $x_{1}^{2}+x_{2}^{2}+x_{3}^{2}=|\bar{x}|^{2}\geq |\bar{x}|_{0}^{2}\}$ outside a a compact set in $\Sigma$ such that
\begin{enumerate}

\item $\ln N=O^{2}(\frac{1}{|\bar{x}|})$ and $\ {\tt g}_{ij}-\delta_{ij}=O^{2}(\frac{1}{|\bar{x}|^{2}})$; where we use the notation $\phi(\bar{x})=O^{2}(f(|\bar{x}|))$ to mean that for some positive numbers $c_{1}$, $c_{2}$ and $c_{3}$ we have 
\ben
|\phi |\leq c_{1}|f(|\bar{x}|)|,\ |\partial_{i} \phi |\leq c_{2} |\partial_{|\bar{x}|} f(|\bar{x}|)|\ {\rm and}\ |\partial_{i}\partial_{j} \phi|\leq c_{3}|\partial^{2}_{|\bar{x}|} f(|\bar{x}|)|.
\een  

\item The second derivatives of $\ln N$ and $\tg_{ij}-\delta_{ij}$ have bounded $C^{\alpha}$-norm (defined with respect to the coordinate system $\{\bar{x}\}$) bounded; namely if $\phi=\partial_{k}\partial_{l} \ln N$ or $\phi=\partial_{k}\partial_{l} (\tg_{ij}-\delta_{ij})$ for all $1\leq k,l,i,j\leq 3$ then
\ben
\|\phi\|_{C^{\alpha}}=\sup_{|\bar{x}-\bar{x}'|\leq 1}\frac{|\phi(\bar{x})-\phi(\bar{x}')|}{|\bar{x}-\bar{x}'|^{\alpha}}<\infty.
\een   

\end{enumerate}

\begin{Proposition}\label{PBS}{\rm{(\bf Beig-Simon\ \cite{MR608121})}} 
Let $(\Sigma,\tg,\ln N)$ be an asymptotically flat static solution. Then, there is a coordinate system $\{\bar{x}=(x_{1},x_{2},x_{3}),\ |\bar{x}|\geq |\bar{x}|_{1}\}$ (not necessarily equal to the one defining asymptotic flatness), such that
\be\label{LNE}
\ln N^{2}=-\frac{2m}{|\bar{x}|}+O^{2}(\frac{1}{|\bar{x}|^{3}}),
\ee
\be
\tg_{ij}=\delta_{ij}-\frac{m^{2}}{|\bar{x}|^{4}}(\delta_{ij}|\bar{x}|^{2}-x_{i}x_{j})+O^{2}(\frac{1}{|\bar{x}|^{3}}).
\ee

\n where $|\bar{x}|^{2}=x_{1}^{2}+x_{2}^{2}+x_{3}^{2}$ and $m$ is the ADM mass of the solution.
\end{Proposition}

\n Note that the remainders are $O^{2}(1/|\bar{x}|^{3})$ in particular $\ln N$ has zero dipole moment. This fact will be important later. Note too that $|\bar{x}|^{2}d\Omega^{2}=|\bar{x}|^{2}(d\theta^{2}+\sin^{2}\theta d\varphi^{2})=(\delta_{ij}-(x_{i}x_{j})/|\bar{x}|^{2})dx_{i}dx_{j}$ therefore we have 
\ben
\tg=\delta_{ij}dx^{i}dx^{j}-m^{2}d\Omega^{2}+O^{2}(\frac{1}{|\bar{x}|^{3}})=(d|\bar{x}|)^{2}+(|\bar{x}|^{2}-m^{2})d\Omega^{2}+O^{2}(\frac{1}{|\bar{x}|^{3}}).
\een

\n To make contact with the representation (\ref{SchP}) of the Schwarzschild solution proceed as follows. Let $(|\bar{x}|,\theta,\varphi)$ be the spherical coordinate system associated to the coordinate system $\{\bar{x}\}$. Make the change of variables $(|\bar{x}|,\theta,\varphi)\rightarrow (r,\theta,\varphi)$ with $r=|\bar{x}|+m$. Then, for the metric $\tg$, we obtain
\ben
\tg = dr^{2}+r^{2}(1-\frac{2m}{r})d\Omega^{2}+O^{2}(\frac{1}{r^{3}})=\tg_{S}+O^{2}(\frac{1}{r^{3}}).
\een

\n For the Lapse $N$ instead, we obtain the following expansion. From (\ref{LNE}) we have
\ben 
N^{2}=1-\frac{2m}{|\bar{x}|}+\frac{2m^{2}}{|\bar{x}|^{2}}+O^{2}(\frac{1}{|\bar{x}|^{3}}).
\een

\n Now use 
\ben
\frac{1}{|\bar{x}|}=\frac{1}{r-m}=\frac{1}{r}+\frac{m}{r^{2}}+\frac{m^{2}}{r^{3}}+O^{2}(\frac{1}{r^{4}}).
\een

\n to get
\ben
N^{2}=1-\frac{2m}{r}+O^{2}(\frac{1}{r^{3}}).
\een

We can thus rephrase the Proposition \ref{PBS} in the following form 

\begin{Proposition}\label{PBS2} 
Let $(\Sigma,\tg,\ln N)$ be an asymptotically flat static solution. Then, there is a coordinate system $\{\bar{x}=(x_{1},x_{2},x_{3}),\ (x_{1}^{2}+x_{2}^{2}+x_{3}^{2})^{\frac{1}{2}}=r\geq r_{1}\}$ (not necessarily equal to the one defining asymptotic flatness), such that
\be\label{LNE2}
N^{2}=1-\frac{2m}{r}+O^{2}(\frac{1}{r^{3}}),
\ee
\be
\tg=dr^{2}+r^{2}(1-\frac{2m}{r})d\Omega^{2}+O^{2}(\frac{1}{r^{3}}).
\ee

\n where $m$ is the ADM mass of the solution.
\end{Proposition}

The following Proposition on the asymptotic of the mean curvatures of the coordinate spheres $S_{r}=\{p/r(p)=r\}$ is now direct.

\begin{Proposition}\label{PMC}
Let $(\Sigma,g,N)$ be an asymptotically flat static solution and consider a coordinate system as in Proposition \ref{PBS2}. Then, the mean curvature $\theta_{r}$ of the level surfaces $S_{r}=\{p/r(p)=r\}$ satisfy, at every point in $S_{r}$, the estimate
\be
\theta_{r}=\frac{2}{r}+\frac{2m}{r^{2}}+O(\frac{1}{r^{3}}). 
\ee
\end{Proposition}

\subsection{The coordinate-distance lag.}\label{SCDL}

Let $(\Sigma,\tg, \ln N)$ be an asymptotically flat static solution with regular and connected horizon $H$. We would like first to introduce the {\it distance function} to $H$, the definition of which is more or less evident. We will follow the Notation \ref{NOT1}. 

Let $p\in \Sigma\setminus H$ and let $\{\Gamma_{i}\}_{i=1}^{i=\infty}$ be a strictly decreasing sequence such that, $\Gamma_{i}\leq \Gamma_{0}$, $\lim \Gamma_{i}=0$ and $p\notin \Omega_{H_{1},H}$. We note that if $j>i$ then 
\ben
dist(p,H_{\Gamma_{i}})<dist(p,H_{\Gamma_{j}}),
\een

\n and we have 
\be\label{INEQ}
dist(p,H_{\Gamma_{i}})\leq dist(p,H_{\Gamma_{j}})\leq dist(p,H_{\Gamma_{i}})+diam(\Omega_{H_{\Gamma_{j}},H_{\Gamma_{i}}}),
\ee

\n where the diameter of $\Omega_{H_{\Gamma_{j}},H_{\Gamma_{i}}}$, $diam(\Omega_{H_{\Gamma_{j}},H_{\Gamma_{i}}})$, tends to zero as $i (<j)\rightarrow \infty$. Denote $s_{\Gamma}(p):=dist(p, H_{\Gamma})$. The inequality (\ref{INEQ}) shows that 
\ben
s(p):=\lim_{i\rightarrow \infty} s_{\Gamma_{i}}(p),
\een

\n for any sequence $\{\Gamma_{i}\}$ as above, is well defined and independent on $\{\Gamma_{i}\}$. We thus define the distance from $p$ to $H$ in that way. Note that given a point $p$ in $\Sigma\setminus H$ one can always construct a length minimizing geodesic from $p$ to $H$ by taking the limit of length minimizing geodesics from $p$ to $H_{\Gamma_{i}}$. This fact will be used later without further mention.

Now consider the Schwazschild solution $\tbg_{S}=dr^{2}+(1-\frac{2m}{r})r^{2}d\Omega^{2}$ and consider a ray $\gamma (r)=(r,\theta_{0},\varphi_{0})$, $r\in [2m,\infty)$, which is, naturally, length minimizing between any two of its points. Let $\st(\gamma(r))$ be the length of $\gamma$ between $r=2m$ and $r$. Then $\st(\gamma(r))=r-2m$ and therefore the limit $lim_{r\rightarrow \infty} \st(\gamma(r))-r+2m=0$. Now consider a ray $\gamma(\tau)$ on an (another) asymptotically flat static solution with regular and connected horizon $H$, joining $H$ to infinity. Then in this different scenario, instead, the limit $\lim s(\gamma(\tau))-r(\gamma(\tau))+2m$ may be different from zero. We advocate now to define {\it the coordinate distance lag} measuring precisely this a priori mismatch. 
\begin{Definition} \label{CDLD}
Let $(\Sigma, \tg, \ln N)$ be an asymptotically flat static solution with connected and regular horizon. Let $\{\bar{x}=(x_{1},x_{2},x_{3}),\ |\bar{x}|=r\geq r_{1}\}$ be a coordinate system as in Proposition \ref{PBS2}. Let $\{p_{i}\}$ be a diverging sequence of points (i.e. $s(p_{i})\rightarrow \infty$) (lying inside the range of $\{\bar{x}\}$). Then, the coordinate distance lag, $\bar{\delta}$, associated to the sequence $\{p_{i}\}$ is defined as 
\ben
\bar{\delta}=\limsup_{i\rightarrow \infty}\  \st(p_{i})-r(p_{i})+2m.
\een
\end{Definition}

\vs 
\n Note that coordinate-distance lags are always zero in the Schwarschild solution. From the next Proposition it will follow that coordinate-distance lags are always finite.
	
\begin{Proposition}\label{CDLF} Let $(\Sigma,\tg,\ln N)$ be an asymptotically flat static solution with connected and regular horizon $H$. Let $\{\bar{x}=(x_{1},x_{2},x_{3}),\ |\bar{x}|=r\geq r_{1}\}$ be a coordinate system as in Proposition \ref{PBS2}. Then there are finite $c_{1}>c_{2}$, depending on $(\Sigma,\tg,\ln N)$, with the following property: for every divergent sequence of points $\{p_{i}\}$ (lying inside the range of $\{\bar{x}\}$) we have
\be\label{CDL}
\st(p_{i})-c_{2}\leq r(p_{i})\leq \st(p_{i})-c_{1}.
\ee
\end{Proposition} 

\n {\it Proof:}

\vs
	We start showing the first inequality in equation (\ref{CDL}). Let us first consider $r_{2}$ such that for every $\bar{x}$ such that $r(|x|)\geq r_{2}$ and a tangent vector $v$ at $\bar{x}$ we have
\ben
\frac{|R(v,v)|}{|\tg_{S}(v,v)|}\leq \frac{R_{0}}{r^{3}}\leq 1,
\een
 
\n where $R$ is the remainder tensor $R:=\tg-\tg_{S}$, $\tg_{S}$ is the Schwarsdchild metric (\ref{SchP}) and $R_{0}$ is a positive constant. It is clear that we do not loose anything in assuming that $r_{2}=r_{1}$.  

Let $d_{0}=sup_{q\in S_{r_{2}}}\{dist(q,H)\}$ and for each $i\geq 0$ consider the curve $\alpha(r)=(r,\theta(p_{i})),\varphi(p_{i}))$ starting at $S_{r_{2}}$ and ending at $p_{i}$ (namely the range of $r$ is $[r_{2},r(p_{i})]$. We will make use of the inequality 
\be\label{SQE}
\sqrt{1+x}\leq 1+|x|,\ {\rm if}\ |x|<1,  
\ee

\n to estimate the distance $\st(p_{i})$ from above. We have
\be\label{SEA}
\st(p_{i})\leq d_{0}+\int_{r_{2}}^{r(p_{i})} \sqrt{\tg_{S}(\alpha',\alpha')+R(\alpha',\alpha')}dr.
\ee

\n As the integration is on $[r_{2},r(p_{i})]$ we have, by the definition of $r_{2}$, $|R(\alpha',\alpha')|/|\tg_{S}(\alpha',\alpha')|\leq R_{0}/r^{3}\leq 1$ (note that $\alpha'=\partial_{r}$). Thus by inequality (\ref{SQE}) we have
\ben 
\sqrt{\tg_{S}(\alpha',\alpha')+R(\alpha',\alpha')}\leq \sqrt{\tg(\alpha',\alpha')}+\frac{R_{0}}{r^{3}}.
\een

\n Putting this into equation (\ref{SEA}) and integrating we have
\ben
\st(p_{i})\leq r(p_{i}) + (d_{0}+\frac{R_{0}}{2r_{2}^{2}}-r_{2}).
\een

\n This proves the first inequality. 

To show the second inequality on the right hand side of equation (\ref{CDL}) we proceed as follows. Consider now an arbitrary curve $\alpha(\tau)$ joining $S_{r_{2}}$ to $p_{i}$, lying inside the region enclosed by $S_{r_{2}}$ and $S_{r(p_{i})}$ and parameterized by the arc length, with respect to $\tg_{S}$, $\tau$. Then, for the length of $\alpha$, $l(\alpha)$, we have 
\ben
l(\alpha)=\int \sqrt{\tg_{S}(\alpha',\alpha')+R(\alpha',\alpha')}d\tau.
\een

\n We are going to make use of the inequality 
\be\label{SQE2}
1-|x|\leq \sqrt{1+x},\ {\rm if}\ |x|\leq 1.
\ee

\n Note that because $\tg_{S}(\alpha',\alpha')=1$ we have $|R(\alpha',\alpha')|\leq R_{0}/r^{3}$. Therefore, from the inequality (\ref{SQE2}) we have 
\be\label{SI}
l(\alpha)\geq \int (1-\frac{R_{0}}{r^{3}})d\tau.
\ee

\n Now note that $|dr/d\tau|\leq 1$. To see this consider an arbitrary parameterization of $\alpha$ by, say $t$. Then $d\tau/dt=\sqrt{\tg_{S}(\partial_{t}\alpha,\partial_{t}\alpha)}\geq |dr/dt|$. 
Thus, noting that the integrand in equation (\ref{SI}) is positive, we can write
\ben
l(\alpha)\geq \int (1-\frac{R_{0}}{r^{3}})d\tau\geq \int (1-\frac{R_{0}}{r^{3}})|\frac{dr}{d\tau}|d\tau\geq \int (1-\frac{R_{0}}{r^{3}})\frac{dr}{d\tau}d\tau.
\een
	
\n Integrating we get
\be\label{FF}
l(\alpha)\geq r_{i}-r_{2}-\frac{R_{0}}{2r_{2}^{2}}.
\ee

\n Now clearly we have $\st(p_{i})$ is greater or equal than the infimum of the lengths of all the curves $\alpha$ joining $p_{i}$ to $S_{r_{2}}$ and lying inside the region enclosed by $S_{r_{2}}$ and $S_{r(p_{i})}$. By the estimation in equation (\ref{FF}) above we have thus 
\ben
\st(p_{i})\geq r(p_{i})-(r_{2}+\frac{R_{0}}{2r_{2}}).
\een

\n which proves the inequality on the right hand side of equation (\ref{CDL}).\ep 	
 	
\begin{Corollary}
Let $(\Sigma,\tg,\ln N)$ be an asymptotically flat static solution with connected and regular horizon $H$. Let $\{\bar{x}=(x_{1},x_{2},x_{3}),\ |\bar{x}|=r\geq r_{1}\}$ be a coordinate system as in Proposition \ref{PBS2}. There are $c_{1}>c_{2}$ depending on $(\Sigma,\tg,\ln N)$ with the following property: for every diverging sequence of points $\{p_{i}\}$ (lying inside the range of $\{\bar{x}\}$) we have
\ben
c_{2}\leq \bar{\delta}(\{p_{i}\})\leq c_{1}.
\een
\end{Corollary} 

\subsection{Distance comparison.}\label{SDC}
	
Consider an asymptotically flat static solution with regular and connected horizon, $(\Sigma,\tg,\ln N)$. Let $\st(p)=dist(p,H)$. If the the solution $(\Sigma,\tg,\ln N)$ were the Schwarzschild solution then we would have 
\ben
\st(p)=r(p)-2m=\frac{2m}{1-N(p)^{2}}-2m.
\een

\n As it turns out, given an arbitrary solution $(\Sigma,\tg,\ln N)$, the function $\hat{\st}$ defined exactly by 
\ben
\hat{\st}(p):=\frac{2m}{1-N(p)^{2}}-2m,
\een

\n provides, via a {\it comparison of Laplacians}, a lower bound for the distance function $\st$. The next Proposition computes the expression of the Laplacian of $\hat{\st}$. 
\begin{Proposition} Let $(\Sigma,\tg,\ln N)$ be a static solution of the Einstein equations. Then, the Laplacian of $\hat{\st}$ has the following expression
\be\label{SBF}
\Delta \hat{\st}=\frac{2}{\bar{s}+2m}(1+\frac{m}{(\hat{\st}+2m)N^{2}})|\nabla \hat{\st}|^{2}.
\ee
\end{Proposition}
	
\vs
\n {\it Proof:}

\vs
Note first the identities 
\be\label{SBI}
\hat{\st} =2m\frac{N^{2}}{1-N^{2}},
\ee
\be\label{SBI2}
N^{2}=\frac{\hat{\st}}{\hat{\st}+2m},\ N^{2}+1=2\frac{\hat{\st}+m}{\hat{\st}+2m}. 
\ee

\n We calculate
\ben
\nabla \frac{1}{1-N^{2}}=2\frac{N\nabla N}{(1-N^{2})^{2}}=2\frac{N^{2}}{(1-N^{2})^{2}}\nabla \ln N.
\een
	
\n Next we compute the divergence of this expression to get
\ben
\Delta \frac{1}{1-N^{2}}=4\frac{|\nabla N|^{2}}{(1-N^{2})^{2}} + 8 \frac{N^{2}|\nabla N|^{2}}{(1-N^{2})^{3}}=4\frac{|\nabla N|^{2}}{(1-N^{2})^{3}}(1+N^{2}),
\een

\n where we have used the fact that $\Delta \ln N=0$. This expression is equal to
\ben
\Delta \frac{1}{1-N^{2}}=|\nabla \frac{1}{1-N^{2}}|^{2}(1+N^{2})(\frac{1-N^{2}}{N^{2}}).
\een

\n After inserting back the coefficient $2m$ and using the identity (\ref{SBI}) we get
\ben
\Delta \hat{\st}=\frac{1+N^{2}}{\hat{\st}}|\nabla \hat{\st}|^{2}.
\een

\n Finally, using the identity (\ref{SBI2}) we have
\ben
\frac{N^{2}+1}{\hat{\st}}=2\frac{\hat{\st}+m}{\hat{\st}+2m}\frac{1}{\hat{\st}}=\frac{2}{\hat{\st}+2m}(1+\frac{m}{(\hat{\st}+2m)N^{2}}).
\een
\ep	

The asymptotic behavior of $\hat{\st}(p)$, when $r(p)\rightarrow \infty$ is deduced from Proposition \ref{PBS2} and we have 
\be\label{SBA}
\hat{\st}(p)=\frac{2m}{\frac{2m}{r(p)}+O(\frac{1}{r(p)^{3}})}-2m=r(p)-2m+O(\frac{1}{r(p)}),
\ee

\n if $r(p)$ is big enough. This asymptotic expression will be important and will be used many times later. 	
	
The reason why we have expressed the Laplacian of $\hat{\st}$ in the form (\ref{SBF}) was to make it comparable with the Laplacian of $\st$, that satisfies the inequality 
\be\label{SL}
\Delta \st\leq \frac{2}{\st+2Pm}(1+\frac{Pm}{(\st +2Pm)N^{2}})|\nabla \st|^{2}.
\ee

\n in a certain {\it barer sense} as is explained in Proposition \ref{LSH}. In the equation above $P$ is equal to the expression
\ben
P=\frac{A}{16\pi m^{2}},
\een

\n and will be called {\it the Penrose quotient}. Note that the Penrose inequality $A\leq 16\pi m^{2}$ holds iff $P\leq 1$. Note too that wherever $s$ is smooth we have $|\nabla s|^{2}=1$. We have included such factor in (\ref{SL}) to make the comparison to (\ref{SBF}) more evident.

The fact that the inequality (\ref{SL}) holds in a barer sense will allow us to assume, when comparing $\st$ to $\hat{\st}$, that $\st$ is a smooth function. This fact will be further explained in Theorem \ref{DC}. We now introduce a Proposition describing the {\it sense} in which inequality (\ref{SL}) holds.

\begin{Proposition}\label{LSH}
Let $(\Sigma,\tg,\ln N)$ be an asymptotically flat static solution with regular and connected horizon. Let $\{p_{i}\}_{i=1}^{i=\infty}$ be a sequence of points in $\Sigma$ converging to $p$ in $\Sigma\setminus H$. Let $\{\Gamma_{i}\}_{i=1}^{i=\infty}$ be a sequence such that $\lim_{i\rightarrow 0} \Gamma_{i}\downarrow 0$, $\Gamma_{1}\leq \Gamma_{0}$ with $\Gamma_{0}$ regular (Notation \ref{NOT1}) and $\{p_{i},i=1,\ldots,i=\infty\}\subset \Sigma\setminus \Omega_{H,H_{\Gamma_{0}}}$. Consider the sequence of distance functions $\{\st_{\Gamma_{i}}(p)=dist(p,H_{\Gamma_{i}})\}_{i=1}^{i=\infty}$. Then, there is sequence of continuous functions $\tilde{\st}_{\Gamma_{i}}$ such that for each $\Gamma_{i}$:
\begin{enumerate}
\item $\tilde{\st}_{\Gamma_{i}}$ is defined on the domain $\Sigma\setminus \Omega_{H,H_{\Gamma_{i}}}$,

\item $\tilde{\st}_{\Gamma_{i}}$ is smooth at $p_{i}$, 

\item $\tilde{\st}_{\Gamma_{i}}\geq \st_{\Gamma_{i}}$, $\tilde{\st}_{\Gamma_{i}}(p_{i})= \st_{\Gamma_{i}}(p_{i})$ and $|\nabla \tilde{\st}|^{2}(p_{i})=1$.

\item
\ben
\Delta \tilde{\st}_{\Gamma_{i}}(p_{i})\leq 2\frac{1}{\tilde{\st}_{\Gamma_{i}}(p_{i})+\tilde{a}_{i}}(1+\frac{\tilde{a}_{i}}{2(\tilde{\st}_{\Gamma_{i}}(p_{i})+\tilde{a}_{i}) N^{2}(p_{i})})|\nabla \tilde{\st}_{\Gamma_{i}}|^{2}(p_{i}),
\een
\n where $\{\tilde{a}_{i}\}$ is a sequence such that $\lim_{i\rightarrow \infty} \tilde{a}_{i}=2mP$.
\item  Moreover, $\{\tilde{\st}_{\Gamma_{i}}\}$ converges uniformly in $C^{0}$ to $\st(p)=dist(p,H)$ in the sense that 
\ben
\lim_{i\rightarrow \infty} \sup_{q\in \Sigma\setminus \Omega_{H,H_{\Gamma_{i}}}}|\tilde{\st}_{\Gamma_{i}}(q)-\st(q)|=0.
\een
\end{enumerate}
\end{Proposition}
The proof of this Proposition will be a direct consequence of the following Proposition in Riemannian geometry. We will use the following notation and terminology. 
\begin{Notation}\label{NOT2}
Let $(\Sigma, g)$ be a complete Riemannian manifold with non-empty and connected boundary $\partial \Sigma$. The {\it inner-normal bundle} ${\mathcal{N}}(\partial \Sigma)$ of $\Sigma$ at $\partial \Sigma$ is defined as the set of vectors $v(q)$, normal to $\partial \Sigma$ at $q$, and pointing inwards to $\Sigma$. We will consider the exponential map $exp:{\mathcal{N}}(\partial \Sigma)\rightarrow \Sigma$ such that to every $v(q)\in {\mathcal{N}}(\partial \Sigma)$ assigns the end point of the geodesic segment of length $|v(q)|$ that start at $q$ with velocity $v(q)/|v(q)|$. 
\end{Notation}

\begin{Proposition}\label{LLSH}
Let $(\Sigma, g)$ be a complete Riemannian three-manifold, not necessarily compact. Let ${\mathcal{S}_{1}}$ be an immersed smooth surface separating $\Sigma$ into two connected (open) components $\Sigma_{1}$ and $\Sigma_{2}$. Let $p$ be a point in $\Sigma_{1}$ and $\gamma_{q,p}$ be a geodesic segment minimizing the distance between $p$ and $\partial \Sigma_{1}={\mathcal{S}_{1}}$, starting at $q\in \partial \Sigma_{1}$ and ending at $p$. We can write $\gamma_{q,p}(\tau)=exp(\tau v(q))$, $\tau\in[0,1]$, with $v(q)=l(\gamma_{q,p})n(q)$  where $n(q)$ is the inward unit-normal vector to $\partial \Sigma_{1}$ at $q$. If the differential of the exponential map $exp:{\mathcal{N}}(\partial \Sigma_{1})\rightarrow \Sigma_{1}$ is not injective at $v(q)$, then for every smooth surface ${\mathcal{S}}_{2}$ immersed in $\Sigma_{1}\cup {\mathcal{S}}_{1}$ such that 
\begin{enumerate}
\item ${\mathcal{S}}_{2}$ touches ${\mathcal{S}}_{1}$ only at $q$,

\item The second fundamental forms $\Theta_{1}(q)$ and $\Theta_{2}(q)$ of ${\mathcal{S}}_{1}$ and ${\mathcal{S}}_{2}$ (respectively) at $q$ and defined with respect to $n(q)$ satisfy
\ben
\Theta_{2}(q)>\Theta_{1}(q).
\een
\end{enumerate}

\n we have, 
\begin{enumerate}
\item $\gamma_{q,p}$ is the only geodesic segment minimizing the distance between $p$ and ${\mathcal{S}}_{2}$,

\item The exponential map $exp:{\mathcal{N}}(\partial \tilde{\Sigma}_{1})\rightarrow \tilde{\Sigma}_{1}$ is injective at $v(q)$, where $\tilde{\Sigma}_{1}$ is the connected component of $\Sigma\setminus {\mathcal{S}}_{2}$ containing ${\mathcal{S}}_{1}$.
\end{enumerate}
\end{Proposition}

\n {\it Proof:}

\vs
First it is clear that $\gamma_{p,q}$ is the only geodesic segment minimizing the distance between $p$ and ${\mathcal{S}}_{2}$ for ${\mathcal{S}}_{2}$ touches ${\mathcal{S}}_{1}$ only at $q$. This proves the first {\it item} of the claim. 

To prove the second suppose on the contrary that the exponential map $exp:{\mathcal{N}}(\partial \tilde{\Sigma}_{1})\rightarrow \tilde{\Sigma}_{1}$ is not injective at $v(q)$. Then there is a curve $w(\lambda)$, $\lambda\in [0,\lambda_{1}]$ of vectors in ${\mathcal{N}}(\tilde{\Sigma})$ of norm (for all $\lambda$) equal to $l(\gamma_{p,q})$, such that $w(0)=v(q)$ and such that $d\ exp (w'(0))=0$. Therefore $J(s)=d\ exp(\frac{s}{l(\gamma_{p,q})}w'(0)$ is a Jacobi field such that $J(s)\neq 0$ for any $s\in [0,l(\gamma_{p,q}))$. Let $\alpha(s,\lambda)$, $(s,\lambda)\in [0,l(\gamma_{p,q})]\times [0,\lambda_{1}]$ be a smooth one-parameter family of curves such that $\partial_{\lambda} \alpha (s,0)=J(s)$ and such that $\partial_{s}\alpha (0,\lambda)\in {\mathcal{N}}(\tilde{\Sigma}_{1})$. Then because $J(s)$ is a Jacobi field we have that the second variation of the length of the curves $\alpha_{\lambda}(s)=\alpha(s,\lambda)$ (variation with respect to $\lambda$) is equal to zero\footnote{Although it is a standard fact in Riemannian geometry, the reader can check this fact in pages 227-228 of \cite{MR757180}. The proof there is for Jacobi fields vanishing at the two extreme points, but it is simply adapted to this situation as well.}. On the other hand consider the curves $\bar{\alpha}(s,\lambda)=\alpha(s,\lambda)$, with $(s,\lambda)\in [0,s(\lambda)]\times [0,l(\gamma_{p,q})]$ where the point $\alpha(s(\lambda),\lambda)$ is the intersection of $\alpha(s,\lambda)$ (a curve as a function of $s$) and ${\mathcal{S}}_{1}$. Now, because of the conditon in {\it item 2}, $\Theta_{2}(q)>\Theta_{1}(q)$, the second variation (with respect to $\lambda$) of $\bar{\alpha}$ is positive. Thus the second variation (with respect to $\lambda$) of the length of the curves $\tilde{\alpha}(s,\lambda)=\alpha(s,\lambda)$, $(s,\lambda)\in [s(\lambda),l(\gamma_{p,q})]\times [0,\lambda_{1}]$ is negative, which is a contradiction as $\gamma_{p,q}$ is length minimizing between $p$ and ${\mathcal{S}}_{1}$.\ep      

\vs
\n {\it Proof ({\rm{\bf of Proposition \ref{LSH}}})}:

\vs
Let $\gamma_{p_{i},q_{i}}$, $q_{i}\in H_{\Gamma_{i}}$ be a length minimizing geodesic joining $p_{i}$ and $H_{\Gamma_{i}}$. Suppose first that $\st_{\Gamma_{i}}$ is smooth at $p_{i}$ for each $i$. Then we claim that taking $\tilde{\st}_{\Gamma_{i}}=\st_{\Gamma_{i}}$ is enough. It is clear that the {\it items 1,2,3} and {\it 5} of the claim are satisfied with this choice.  We need therefore to check that there is sequence $\tilde{a}_{i}$ for which the equation in {\it item 4} is satisfied and $\lim_{i\rightarrow \infty} \tilde{a}_{i}=2mP$. For this we are going to use the monotonicity, for every $a$ of ${\mathcal{M}}={\mathcal{M}}_{a}$ an over $\gamma_{p_{i},q_{i}}$, and then we will chose $a$ conveniently (which will be our choice of $\tilde{a}_{i}$). Of course ${\mathcal{M}}$ is defined, for each $i$, for the congruences ${\mathcal{F}}_{i}$ of length minimizing geodesics segments to $H_{\Gamma_{i}}$. Thus we have
\ben
\frac{\theta(p_{i})}{2}(\st_{\Gamma_{i}}(p_{i})+a)^{2}N^{2}(p_{i})-(\st_{\Gamma_{i}}(p_{i})+a)N^{2}(p_{i})={\mathcal{M}}_{a}(p_{i})\leq {\mathcal{M}}_{a}(q_{i}).
\een 

\n Solving for $\theta(p_{i})=\Delta \st_{\Gamma_{i}} (p_{i})$ we get
\ben
\Delta \st_{\Gamma_{i}}(p_{i})\leq \frac{2}{(\st_{\Gamma_{i}}(p_{i})+a)}(1+\frac{{\mathcal{M}}_{\Gamma_{i}}(q_{i})}{(\st_{\Gamma_{i}}(p_{i})+a)N^{2}(p_{i})}).
\een

\n We need now to show that we can chose $a$ for each $i$ (thus having $a=\tilde{a}_{i}$) in such a way that    
${\mathcal{M}}_{\Gamma_{i}}(q_{i})\leq \tilde{a}_{i}/2$. Therefore we need to have
\ben
{\mathcal{M}}_{\Gamma_{i}}(q_{i})=\frac{\theta_{\Gamma_{i}}(q_{i})}{2}a^{2}N^{2}(q_{i})-aN(q_{i})^{2}\leq \frac{a}{2}.
\een

\n Thus we chose
\be\label{ADEF}
a= \sup_{q\in H_{\Gamma_{i}}} \{\frac{2(\frac{1}{2}+N(q)^{2})}{\theta(q)N^{2}(q)}\}.
\ee

\n Now, the numerator tends to one and the denominator, because of equation (\ref{MH}), tends to $2|\nabla N|_{H}=8\pi m/A=1/(2mP)$. The claim in this case follows. 

If on the contrary the functions $\st_{\Gamma_{i}}$ are not smooth at $p_{i}$, then we know by Proposition \ref{LLSH} that the distance functions $\tilde{\st}_{\Gamma_{i}}$ to a hypersurface $\tilde{H}_{\Gamma_{i}}$ included in $\Omega_{H,H_{\Gamma_{i}}}$ will be smooth at $p_{i}$ provided they touch $H_{\Gamma_{i}}$ only at $q_{i}$ and have strictly grater second fundamental form at $q_{i}$. Besides these last two conditions nothing else is required on the hypersurfaces $\tilde{H}_{\Gamma_{i}}$ for $\tilde{\st}_{\Gamma_{i}}$ to be smooth at $p_{i}$. Thus, it is clear that if we chose the hypersurfaces $\tilde{H}_{\Gamma_{i}}$ close enough to $H_{\Gamma_{i}}$ (but satisfying the two requirements) and $\tilde{a}_{i}$ using the same formula as in equation \ref{ADEF} (but with $q$ varying on $\tilde{H}_{\Gamma_{i}}$) then $\tilde{\st}_{\Gamma_{i}}$ will satisfy {\it items 1- 5} of the claim. \ep

\begin{Theorem}\label{DC}{\rm{\bf (Distance comparison).}} Let $(\Sigma,\tg,\ln N)$ be an asymptotically flat static solution with regular and connected horizon. Then we have 
\be\label{MDC}
\frac{2m}{1-N^{2}(p)}-2m=\hat{\st}(p)\leq \max\{1,\frac{1}{P}\}\st(p)=\max\{1,\frac{16 \pi m^{2}}{A}\} dist(p,H),
\ee

\n for all $p$ in $\Sigma$, where $P$ is the Penrose quotient. Moreover 
\ben
\lim_{\st(p)\rightarrow \infty} \frac{\hat{\st}(p)}{\st(p)}=1,\ {\rm and}\ \lim_{\st(p)\rightarrow 0} \frac{\hat{\st}(p)}{\st(p)}=\frac{1}{P},
\een
\end{Theorem}    

\n {\it Proof:}

\vs
We will consider the quotient $\hat{\st}/\st$ as a function on $\Sigma\setminus H$. Let us first find the boundary conditions, namely $\lim \hat{\st}(p)/\st(p)$ when $\st(p)\rightarrow \infty$ and $\st(p)\rightarrow 0$ (at infinity and at the horizon respectively). From Proposition \ref{CDLF} and the estimation (\ref{SBA}) we deduce 
\ben
\lim_{\st(p)\rightarrow \infty} \frac{\hat{\st}(p)}{\st(p)}=1. 
\een  
	
\n To calculate the quotient at the horizon we proceed like this. Consider the congruence of geodesics with respect to $g$, emanating perpendicularly to $H$ and parameterized by the arc length $\tau$ which is measured from the initial point of the geodesic at $H$. Any given coordinate system $\{\bar{x}=(x_{1},x_{2})\}$ on an open set of $H$ can be propagated along the congruence to the level sets of the distance function with respect to $g$, namely the $\tau_{0}$-level sets $\{\tau=\tau_{0}\}$ and we can write
\ben
g=d\tau^{2}+h_{ij}(\bar{x},\tau)dx_{i}dx_{j},
\een

\n and 
\be\label{CCC}
\hat{\st}(\tau,\bar{x})=\frac{2m}{1-N^{2}(\tau,\bar{x})}-2m=2m|\nabla N|^{2}_{H}\tau^{2}+O(\tau^{3}).
\ee

\n We note then that because $H$ is totally geodesic, the second fundamental form is zero and we have 
\ben
\partial_{\tau} h_{ij}(\tau,\bar{x})\bvl_{\tau=0}=0.
\een

\n Thus
\be\label{CCCC}
g=d\tau^{2}+h_{ij}(0,\bar{x})dx_{i}dx_{j}+O(\tau^{2}).
\ee  

\n Combining (\ref{CCC}) and (\ref{CCCC}) we get 
\ben
\tg=N^{2}g=|\nabla N|_{H}^{2}\tau^{2}(d\tau^{2}+h_{ij}(0,\bar{x})dx_{i}dx_{j}) + O(\tau^{3})d\tau^{2}+O(\tau^{4})h_{ij}dx_{i}dx_{j}.
\een

\n From this expression it is simple that if $\{p_{i}\}$ is a sequence in $\Sigma\setminus H$ converging to a point in $H$ we have
\be\label{C4}
\st(p_{i})=|\nabla N|_{H}\frac{\tau(p_{i})^{2}}{2}+O(\tau(p_{i})^{3}).
\ee

\n We can combine (\ref{CCC}) and (\ref{C4}) to conclude that for any sequence $\{p_{i}\}$ in $\Sigma\setminus H$ converging to a point in $H$ we have
\be\label{C5}
\lim \frac{\hat{\st}(p_{i})}{\st(p_{i})}=4m|\nabla N|_{H}.
\ee

\n Now, $|\nabla N|_{H}$ is equal to $4\pi m/A$ as can be seen by integrating $\Delta N=0$ between $S_{r}=\{p/r(p)=r\}$ and $H$ and taking the limit when $r\rightarrow \infty$. With this value of $|\nabla N|_{H}$ we get from (\ref{C5})
\ben
\lim_{\st(p)\rightarrow 0} \frac{\hat{\st}(p)}{\st(p)}=\frac{16\pi m^{2}}{A}=\frac{1}{P}.
\een   

We would like now to compare $\hat{\st}$ to $\st$ using (\ref{SBF}) and (\ref{SL}). For this purpose it is simpler to consider the dimensionless quantities $\hat{u}=\hat{\st}/2m$ and $u=\st/2mP$. In terms of them (\ref{SBF}) and (\ref{SL}) become
\be\label{BSIU0}
\Delta \hat{u}=\frac{2}{\hat{u}+1}(1+\frac{1}{2(\hat{u}+1)N^{2}})|\nabla \hat{u}|^{2},
\ee
\be\label{BSIU}
\Delta u\leq\frac{2}{u+1}(1+\frac{1}{2(u+1)N^{2}})|\nabla u|^{2},
\ee

\n We will consider now the quotient $\phi=\hat{u}/u$ and note that the boundary conditions at $H$ and at infinity become, respectively, $\lim_{\st(p)\rightarrow 0}\hat{u}(p)/u(p)=1$ and $\lim_{\st(p)\rightarrow \infty}\hat{u}(p)/u(p)=P$. If we prove that $\hat{u}/u\leq \max\{1,P\}$ then we will be proving (\ref{MDC}). Thus we will proceed by contradiction and assume that there is a point $\bar{p}\in \Sigma\setminus H$ such that $\hat{u}(\bar{p})>\max\{1,P\}u(\bar{p})$ and that such point is an absolute maximum for $\hat{u}/u$ (note the boundary conditions). We will assume below that the function $\st$ is smooth at $\bar{p}$, or, equivalently that $u$ is smooth at $\bar{p}$. Otherwise use the fact that $\st$ satisfies equation (\ref{SL}) in a barer sense as follows. Replace $\st$ by $\st_{\Gamma}$ for $\Gamma$ sufficiently small in such a way that $\hat{u}/u_{\Gamma}$, with $u_{\Gamma}=\st_{\Gamma}/2mP$ still has a maximum greater than $\max\{1,P\}$, say at $\bar{\bar{p}}$. Then substitute once more $\st_{\Gamma}$ by $\tilde{\st}_{\Gamma}\geq \st_{\Gamma}$ as in Proposition \ref{LSH} and consider thus the quotient $\hat{u}/\tilde{u}_{\Gamma}$, with $\tilde{u}_{\Gamma}=\tilde{\st}_{\Gamma}/2mP$, which still has a maximum greater than $\max\{1,P\}$ at $\bar{\bar{p}}$. If $\Gamma$ is sufficiently small we would reach a contradiction following the same argument as below. 

We compute
\be\label{CLQ}
\Delta \frac{\hat{u}}{u}=\frac{\Delta \hat{u}}{u}-2\frac{<\nabla \hat{u},\nabla u>}{u^{2}}-\frac{\hat{u}}{u^{2}}\Delta u+2\frac{\hat{u}}{u^{3}}|\nabla u|^{2}.
\ee
   
\n Because $\hat{u}/{u}$ reaches an absolute maximum at $\bar{p}$ we have $\nabla (\hat{u}/u|_{\bar{p}})=0$ and thus
\be\label{GEM}
\frac{\nabla \hat{u}}{\hat{u}}\bvl_{\bar{p}}=\frac{\nabla u}{u}\bvl_{\bar{p}},
\ee

\n with $|\nabla u|^{2}(\bar{p})=1/2mP\neq 0$. If we use (\ref{GEM}) in (\ref{CLQ}) we note that the second and fourth terms on the right hand side cancel out at $\bar{p}$. Thus we will get a contradiction of the fact that $\hat{u}/u$ reaches an absolute maximum at $\bar{p}$ if we can prove that the sum of the first and third terms on the right hand side of (\ref{CLQ}) is positive at $\bar{p}$ (the Maximum Principle). We will prove that in what follows.

We compute
\ben
\Delta \frac{\hat{u}}{u}\ \bigg|_{\bar{p}}=\frac{1}{u^{2}(\bar{p})}(u\Delta \hat{u} -\hat{u}\Delta u)\bvl_{\bar{p}}.
\een

\n and using (\ref{BSIU0}) and (\ref{BSIU}) we get the inequality
\ben
\Delta\frac{\hat{u}}{u}\bvl_{\bar{p}}\geq \frac{2}{u^{2}}(\frac{u}{(1+\hat{u})}(1+\frac{1}{2(1+\hat{u})N^{2}}))\frac{\hat{u}^{2}}{u^{2}}|\nabla u|^{2} - \frac{\hat{u}}{(1+u)}(1+\frac{1}{2(1+u)N^{2}})|\nabla u|^{2})\bvl_{\bar{p}}.
\een

\n Thus we would like to prove that 
\be\label{FOR}
\frac{\hat{u}}{1+\hat{u}}(1+\frac{1}{2(1+\hat{u})N^{2}})\bvl_{\bar{p}}>\frac{u}{1+u}(1+\frac{1}{2(1+u)N^{2}})\bvl_{\bar{p}}.
\ee
	
\n Recalling from (\ref{SBI2}) that $N^{2}=\hat{u}/(1+\hat{u})$ and substituting that into (\ref{FOR}) we deduce that we would like to show that
\ben
\frac{\hat{u}}{(1+\hat{u})}(1+\frac{1}{2\hat{u}})\bvl_{\bar{p}}>\frac{u}{1+u}(1+\frac{1+\hat{u}}{2(1+u)\hat{u}})\bvl_{\bar{p}}.
\een 
		
\n We will arrange now this equation in a different form. To this, right hand term $u/(1+u)$ is moved to the left hand side, while the left hand term $1/(2(1+\hat{u}))$ is moved to the right hand side. In this way we obtain a new inequality where the left hand side is
\ben
\frac{\hat{u}}{1+\hat{u}}-\frac{u}{1+u}\bvl_{\bar{p}}=\frac{\hat{u}-u}{(1+u)(1+\hat{u})}\bvl_{\bar{p}},
\een

\n and where the right hand side is 
\ben
\frac{u(1+\hat{u})}{2\hat{u}(1+u)^{2}}-\frac{1}{2(1+\hat{u})}\bvl_{\bar{p}}=\frac{1}{2\hat{u}(1+\hat{u})(1+u)^{2}}(u(1+\hat{u})^{2}-\hat{u}(1+u)^{2})\bvl_{\bar{p}}.
\een

\n This last expression can be further arranged into
\ben
\frac{1}{2\hat{u}(1+\hat{u})(1+u)^{2}}(\hat{u}-u)(\hat{u}u-1)\bvl_{\bar{p}}.
\een

\n Thus combining the results on the left and right hands we conclude that we would like the inequality
\ben
\frac{\hat{u}-u}{(1+u)(1+\hat{u})}\bvl_{\bar{p}}>\frac{1}{2\hat{u}(1+\hat{u})(1+u)^{2}}(\hat{u}-u)(\hat{u}u-1)\bvl_{\bar{p}},
\een

\n to be satisfied. Thus we would like to have
\ben
2(\hat{u}-u)\hat{u}(1+u)\bvl_{\bar{p}}> (\hat{u}-u)(\hat{u}u-1)\bvl_{\bar{p}},
\een

\n but because we are assuming $\hat{u}(\bar{p})>\max\{1,P\}u(\bar{p})\geq u(\bar{p})$ the inequality above is clearly satisfied.\ep 

\subsection{The Penrose inequality.}\label{SPI}

In this section we will prove the Penrose inequality for asymptotically flat static solutions with regular and connected horizon. We start by observing and interesting Corollary to Theorem \ref{DC}.

\begin{Corollary} {\rm (To Theorem \ref{DC})}\label{CDC}
Let $(\Sigma,\tg,\ln N)$ be an asymptotically flat static solution with connected and regular horizon. Suppose that the Penrose inequality does not hold, namely, assume that the Penrose quotient $P=\frac{A}{16\pi m^{2}}$ is greater than one. Then, for any divergent sequence of points $\{p_{i}\}$, the associated coordinate-distance lag is greater or equal than zero, namely $\bar{\delta}(\{p_{i}\})\geq 0$.
\end{Corollary}

\n {\it Proof:}

\vs
If $P>1$ then $\max\{1,\frac{1}{P}\}=1$ and from Theorem \ref{DC} we have then 
\ben
\hat{\st}(p)=\frac{2m}{1-N^{2}(p)}-2m\leq \st(p),\ {\rm for\ all}\ p\in \Sigma.
\een

\n Evaluating this inequality at $\{p_{i}\}$ and using the asymptotic of $\hat{\st}$ described in equation (\ref{SBA}) we get
\ben
 0\leq \st(p_{i})-r(p_{i})+2m+O(\frac{1}{r(p_{i})}),
\een
 
\n Therefore
\ben
0\leq \limsup_{i\rightarrow \infty} \st(p_{i})-r(p_{i})+2m=\bar{\delta}(\{p_{i}\}).
\een

\n as desired.\ep

The following Proposition however shows (in particular) that if the Penrose inequality does not hold then there is a divergent sequence $\{p_{i}\}$ whose coordinate-distance lag is negative, namely $\bar{\delta}(\{p_{i}\})<0$. The two results thus show the Penrose inequality on asymptotically flat static solutions with regular and connected horizon.

\begin{Proposition}\label{PDC}
Let $(\Sigma,\tg,\ln N)$ be an asymptotically flat static solution with regular and connected horizon $H$. Then, there is a divergent sequence $\{p_{i}\}$ such that 
\ben
\bar{\delta}(\{p_{i}\})\leq m(1-P).
\een

\n In particular if $P>1$ then $\bar{\delta}(\{p_{i}\})<0$.
\end{Proposition}

\n {\it Proof:} 

\vs
Let $\{\Gamma_{i}\}_{i=1}^{i=\infty}$ be a sequence such that $\Gamma_{i}\downarrow 0$ (with $\Gamma_{1}\leq \Gamma_{0}$ and $\Gamma_{0}$ regular as in Notation \ref{NOT1}), and let $\{r_{i}\}_{i=1}^{i=\infty}$ be a sequence such that $r_{i}\uparrow \infty$ (and $r_{1}$ as in Proposition \ref{PBS2}). 
Consider the congruence of length minimizing geodesics ${\mathcal{F}}$ emanating perpendicularly to $H_{\Gamma_{i}}$. The geodesic segment, $\gamma_{i}$, minimizing the length between $H_{\Gamma_{i}}$ and $S_{r_{i}}$ is clearly in ${\mathcal{F}}$. Let $p_{i}$ be the point of $\gamma_{i}$ at $S_{r_{i}}$, let $q_{i}$ be the initial point at $H_{\Gamma_{i}}$ and let $v(q_{i})$ the (unit) velocity of $\gamma_{i}$ at $q_{i}$. $\gamma_{i}$ is naturally perpendicular to $S_{r_{i}}$ at $p_{i}$ and to $H_{\Gamma_{i}}$ at $q_{i}$. Consider now the exponential map $exp:{\mathcal{N}}_{i}\rightarrow \Sigma$, where ${\mathcal{N}}_{i}$ is the inner-normal bundle of $\Sigma\setminus \Omega_{H,H_{\Gamma_{i}}}$ at $H_{\Gamma_{i}}$ as in Notation \ref{NOT2}. Assume that the differential of the exponential map is smooth at the point $l(\gamma_{i})v(q_{i})$ in ${\mathcal{N}}_{i}$, if not, work instead with a suitable function $\tilde{\st}_{\Gamma_{i}}$ as in Proposition \ref{LSH}. Note that, in the notation of Proposition \ref{LSH}, we have $l(\gamma_{i})=\st_{\Gamma_{i}}(p_{i})$. Then, there is $\epsilon_{i}$ such that the surface defined by $\bar{S}_{i}=\{exp(l(\gamma_{i})v(q)),\ q\in B_{H_{\Gamma_{i}}}(q_{i},\epsilon_{i})\}$ is smooth. Moreover $\bar{S}_{i}$ is tangent to $S_{r_{i}}$ at $p_{i}$, its mean curvature is equal to the mean curvature  $\theta$ of ${\mathcal{F}}$ restricted to it, and, because $\gamma_{i}$ is length minimizing between $S_{r_{i}}$ and $H_{\Gamma_{i}}$, it lies inside the region enclosed by $H_{\Gamma_{i}}$ and $S_{r_{i}}$. Therefore from the standard comparison of mean curvatures we have
\ben
\theta(p_{i})\geq \theta_{r_{i}}(p_{i}),
\een

\n where $\theta_{r_{i}}$ is the mean curvature of $S_{r_{i}}$. Consider now ${\mathcal{M}}$ with $a=A/8\pi m$ and over $\gamma_{i}$. As ${\mathcal{M}}$ is monotonic we have
\ben
\theta(p_{i})\leq \frac{2}{\st_{\Gamma_{i}}+\frac{A}{8\pi m}} + \frac{2{\mathcal{M}}(q_{i})}{(\st_{\Gamma_{i}}+\frac{A}{8\pi m})^{2}N^{2}(p_{i})}.
\een     

\n Now, to use this equation we need several facts. First, from Proposition \ref{PMC} we have $\theta_{r_{i}}=2/r_{i}+2m/(r_{i}^{2})+O(1/r_{i}^{3})$. Therefore we have
\be\label{THI}
\frac{2}{r_{i}}+\frac{2m}{r_{i}(r_{i}-2m)}+O(1/r_{i}^{3})\leq  \frac{2}{\st_{\Gamma_{i}}+\frac{A}{8\pi m}} + \frac{2{\mathcal{M}}(q_{i})}{(\st_{\Gamma_{i}}+\frac{A}{8\pi m})^{2}N^{2}(p_{i})}.
\ee

\n We can arrange this better as
\be\label{THI2}
\frac{2(\st_{\Gamma_{i}} +\frac{A}{8\pi m} -r_{i})}{r_{i}(\st_{\Gamma_{i}}+\frac{A}{8\pi m})} +\frac{2m}{r_{i}(r_{i}-2m)} - \frac{2{\mathcal{M}}(q_{i})}{(\st_{\Gamma_{i}}+\frac{A}{8\pi m})^{2}N^{2}(p_{i})}\leq O(1/r_{i}^{3}).
\ee

\n Secondly, from Proposition \ref{MH} we have $\lim {\mathcal{M}}(q_{i})=|\nabla N|_{H} (\frac{A}{8\pi m})^{2}=\frac{A}{16\pi m}$. Finally, we have $\lim \st(p_{i})-\st_{\Gamma_{i}}(p_{i})=0$ and from Proposition \ref{CDLF} it is $\lim r_{i}/\st_{\Gamma_{i}}=1$. Multiplying equation (\ref{THI2}) by $\st_{\Gamma_{i}}^{2}$, taking the limsup while using the facts described above gives finally
\ben
\bar{\delta}(\{p_{i}\})=\limsup \st(p_{i})-r_{i}+2m\leq m(1-P).
\een

\n as desired.\ep 

\vs
\n Using Corollary \ref{CDC} and Proposition \ref{PDC} we deduce the Penrose inequality.

\begin{Proposition} {\rm{\bf (The Penrose inequality).}}\label{PIT}
Let $(\Sigma,g,N)$ be an asymptotically flat static solution with a regular and connected horizon $H$. Let $A$ be the area of $H$ and $m$ the ADM mass of the solution. Then
\be\label{PENR}
A\leq 16\pi m^{2}.
\ee
\end{Proposition}
 
\subsection{The opposite Penrose inequality.}\label{OSPI}

In this Section we prove the {\it opposite Penrose inequality} namely that $A\geq 16\pi m^{2}$. The proof will follow after carefully studying the behavior of the quotient $\hat{\st}/\st$ at the  singularity of $\tg$, namely the (unique) horizon $H$, and using then the distance comparison in Theorem \ref{DC}. We will denote by $\kappa$ the Gaussian curvature of the two-metric on $H$ inherited from $g$. 

\begin{Proposition}\label{EMB}
Let $(\Sigma, g, N)$ be an asymptotically flat static solution with regular and connected horizon. Consider a $g$-geodesic $\gamma$ starting perpendicularly from $H$ at $q$, and parameterized with respect to the $g$-arc length of $\gamma$ from $q$, $\tau$. Define $\hat{\hat{\st}}(\gamma(\tau))=\int_{0}^{\tau}N(\gamma(\tau))d\tau$. Then we have
\be\label{LIM0}
\frac{d}{d \hat{\hat{\st}}}\frac{\hat{\st}}{\hat{\hat{\st}}}\bvl_{q}=8m(\frac{4\pi m}{A})^{2} - 2m \kappa\ \bvl_{q}.
\ee
\end{Proposition}

\n {\it Proof:}

\vs
Note that, as is written in the statement of the Proposition, we will work in the natural representation $(\Sigma,g,N)$ of the static solution.  

Now first we note that $d\hat{\hat{\st}}(\tau)/d\tau =N(\alpha(\tau))$. Derivatives with respect to $\tau$ will be denoted by a prima, i.e. $f'(\alpha(\tau))'=d f(\alpha(\tau))/d\tau$. We compute (when $\tau\neq 0$)
\be\label{LIM}
\frac{d}{d\hat{\hat{\st}}} \frac{\hat{\st}}{\hat{\hat{\st}}}=\frac{2m((2\frac{N'}{1-N^{2}}+2\frac{N^{2}N'}{(1-N^{2})^{2}})\hat{\hat{\st}}-2m\frac{N^{2}}{(1-N^{2})})}{\hat{\hat{\st}}^{2}}.
\ee

\n We want to calculate now the limit of this expression when $\tau\rightarrow 0$.  We will separate the right hand side of (\ref{LIM}) into two terms and calculate the limit for each one of them separately. The first limit we will calculate is
\be\label{LIMIN}
\lim_{\tau\rightarrow 0} \frac{4mN^{2} N'}{(1-N^{2})^{2}\hat{\hat{\st}}}=4m|\nabla N|_{H} \lim_{\tau\rightarrow 0} \frac{N^{2}}{\hat{\hat{\st}}},
\ee

\n which arises from the middle term on the right hand side of equation (\ref{LIM}). The right hand side of (\ref{LIMIN}) was obtained using that $N'(\tau)\rightarrow |\nabla N|_{H}$ and $(1-N^{2})^{2}\rightarrow 1$. We calculate now the limit on the right hand side of (\ref{LIMIN}) using L'H${\rm\hat{o}}$pital rule and we have
\ben
\lim_{\tau\rightarrow 0} \frac{N^{2}}{\hat{\hat{\st}}}=\lim_{\tau\rightarrow 0} 2 N'=2|\nabla N|_{H}.
\een

\n Thus we get 
\be\label{1L}
\lim_{\tau\rightarrow 0} \frac{4mN^{2} N'}{(1-N^{2})^{2}\hat{\hat{\st}}}=8m (|\nabla N|^{2}_{H})=8m(\frac{4\pi m}{A})^{2}.
\ee

\n The second limit that we will calculate is
\be\label{2L}
\lim_{\tau\rightarrow 0} \frac{2m}{1-N^{2}} \frac{(2 N'\hat{\hat{\st}}-N^{2})}{\hat{\hat{\st}}^{2}}= 2m \lim_{\tau\rightarrow 0}\frac{(2 N'\hat{\hat{\st}}-N^{2})}{\hat{\hat{\st}}^{2}}.
\ee  

\n which arises from the combination of the first and third term on the right hand side of (\ref{LIM}). Again, to obtain the right hand side of (\ref{2L}), we use the fact that the factor $2m/(1-N^{2})$ would be, in the limit, $2m$. We calculate the limit on the right hand side of (\ref{2L}) by L'H${\rm{\hat{o}}}$pital rule, and obtain
\be\label{3L}
2m \lim_{\tau\rightarrow 0} \frac{ (2N'-2N'+2\hat{\hat{\st}} \frac{N''}{N})}{2\hat{\hat{\st}}}=2m\ Ric(n,n),
\ee

\n where $n=\alpha'(0)$ is the outward $g$-unit normal vector to $H$ at $\alpha(0)$. To obtain the right hand side above we used the static equation (\ref{SE2}), namely $N''(\alpha(0))=Ric(\alpha'(0),\alpha'(0))N(\alpha(0))$ (note that $\alpha(\tau)$ is a $g$-geodesic). 

Recall now the structure equation $2\kappa (q) + |\Theta|^{2}(q)-\theta^{2}(q)=R(q)-2Ric(n(q),n(q))$, where $q$ is a point in $H$. Again, $n$ is the outward $g$-unit normal vector to $H$ at $q$. $\Theta(q)$ and $\theta(q)$ are the second fundamental forms of $H$, calculated using $g$, and evaluated at $q$. For a regular horizon we know that $\Theta=0$, $\theta=0$. $R$ and $Ric$ are the scalar and Ricci curvatures of $g$ respectively. For a static solution $(\Sigma,g,N)$ it is $R=0$ everywhere. $\kappa$, as said above is the Gaussian curvature of $H$ with the two-metric inherited from $g$. Thus, from the structure equation we get that for all $q$ in $H$ we have $\kappa(q)=-Ric(n,n)$. Using this fact in (\ref{3L}) and combining (\ref{3L}) and (\ref{1L}) to complete the limit (\ref{LIM}), we obtain (\ref{LIM0}).\ep

\begin{Proposition}\label{EMB2}
Let $(\Sigma, \tg, N)$ be an asymptotically flat static solution with regular and connected horizon. If there is a point $q$ at $H$ for which
\be\label{GC}
 \kappa(q)< 4(\frac{4\pi m}{A})^{2}.
\ee

\n then there is a point $p$ in $\Sigma\setminus H$ such that $\hat{\st}(p)/\st(p)>1/P$, where $P$ is the Penrose quotient.
\end{Proposition}

\n {\it Proof:}

\vs
Suppose there is a point $q$ in $H$ for which inequality (\ref{GC}) hods. By Proposition \ref{EMB}, there is a $g$-geodesic emanating perpendicularly to $H$ for which 
\ben
\frac{d}{d \hat{\hat{\st}}}\frac{\hat{\st}}{\hat{\hat{\st}}}>0.
\een

\n Also applying L'h\^opital rule we get
\ben
\lim_{\tau\rightarrow 0} \frac{\hat{\st}}{\hat{\hat{\st}}}=\lim_{\tau\rightarrow 0} \frac{\frac{4mNN'}{(1-N^{2})^{2}}}{N}=4m|\nabla N|_{H}=\frac{1}{P}.
\een

\n Therefore we have $\hat{\st}(\gamma(\tau))/\hat{\hat{\st}}(\gamma(\tau))>1/P$ for $\tau$ small. Now we observe that $\hat{\hat{\st}}(\gamma(\tau))\geq \st(\gamma(\tau))$ because $\st$ is the $\tg$-distance function to $H$ and $\hat{\hat{\st}}(\gamma(\tau))$ is the $\tg$-length of $\gamma$ between $\gamma(0)$ and $\gamma(\tau)$. Thus, for $\tau$ small we have
\ben
\frac{\hat{\st}(\gamma(\tau))}{\st(\gamma(\tau))} = \frac{\hat{\st}(\gamma(\tau))}{\hat{\hat{\st}}(\gamma(\tau))}\frac{\hat{\hat{\st}}(\gamma(\tau))}{\st(\gamma(\tau))} \geq 
\frac{\hat{\st}(\gamma(\tau))}{\hat{\hat{\st}}(\gamma(\tau))}>\frac{1}{P}.
\een   
\ep

\begin{Corollary}\label{CORIPI} Let $(\Sigma,g,N)$ be an asymptotically flat static solution with regular and connected horizon $H$. Then, $H$ is homeomorphic to a two-sphere and the inverse Penrose inequality holds, $A\geq 16\pi m^{2}$. Moreover if the Penrose inequality holds, namely $A\leq 16\pi m^{2}$, then $\kappa=4\pi/A$ and the horizon is round.
\end{Corollary}

\n {\it Proof:}

\vs
	By Proposition \ref{EMB2} if there is a point $q$ in $H$ for which $\kappa(q)<4(4\pi m/A)^{2}$ then there is point $p$ in $\Sigma\setminus H$ such that $\hat{s}(p)/s(p)>1/P$ but this contradicts the distance comparison of Theorem \ref{DC}. Therefore $\kappa\geq 4(4\pi m/A)^{2}$ and, by Gauss-Bonnet, $H$ must be homeomorphic to a two sphere. Moreover
\ben
\int_{H}\kappa dA=4\pi\geq 4(\frac{4\pi m}{A})^{2}.
\een

\n Thus
\ben
A\geq 16\pi m^{2},
\een

\n which finishes the first part of the claim. Suppose now that $A\leq 16\pi m^{2}$ then, as $\kappa\geq 4(4\pi m/A)^{2}$ we must have $k=4(4\pi m/A)^{2}=4\pi/A$ which finishes the claim.\ep 

\subsection{The uniqueness of the Schwarzschild solution.}\label{USS}

\subsubsection{Further properties of the coordinate-distance lag.}\label{FPCDL}

The proof of the uniqueness of the Schwarschild solutions does not follows directly in our setting from the equality $A=16\pi m^{2}$. Indeed it is required first to prove that for any divergence sequence $\{p_{i}\}$ the associated coordinate-distance lag $\bar{\delta}(\{p_{i}\})$ is zero. We advocate now to prove this intermediate step. We need two preliminary Propositions. We start showing that $|\nabla \hst|\leq 1$. 

\begin{Proposition}\label{GRADUN} Let $(\Sigma,\tg,\ln N)$ be an asymptotically flat static solution with regular an connected horizon. Then, $|\nabla \hat{\st}|_{\tg}\leq 1$. 
\end{Proposition}

\vs
\n {\it Proof:}

	We observe first that $\lim_{\st(p)\rightarrow \infty}|\nabla \hat{\st}|_{\tg}(p)=1$. But we also have $\lim_{\st(p)\rightarrow 0}|\nabla \hat{\st}|_{\tg}=1$. To see this last claim we compute
\ben
|\nabla \hat{\st}|_{\tg}(p)=\frac{4m}{(1-N^{2}(p))^{2}}|\nabla N(p)|_{g}\rightarrow 4m |\nabla N|_{H}.
\een
	
\n But we already know from Corollary \ref{CORIPI} that $P=1$ and thus $|\nabla N|_{H}=4\pi m/A=1/4m$. The claim follows. 
	
	We show now that there cannot exist a point $p$ in $\Sigma\setminus H$ for which $|\nabla \hat{\st}|(p)>1$. 
We will assume without loss of generality that $m=1$. The assumption simplifies the writing. Define
\ben
\hat{\st}_{\alpha}=\frac{1}{1-N^{2\alpha}}-1,
\een

\n and thus 
\ben
N^{2\alpha}=\frac{\hat{\st}_{\alpha}}{\hat{\st}_{\alpha}+1}.
\een

\n Then we compute
\ben
2\alpha N^{2\alpha-1}\nabla N=\frac{1}{(\sta+1)^{2}}\nabla \sta,
\een

\n and thus
\ben
\frac{\nabla N}{N}=\frac{1}{2\alpha \sta(\sta+1)}\nabla \sta.
\een

\n But $\Delta \ln N=0$ and then $\nabla (1/(\sta(\sta+1))\nabla \sta)=0$ which can be written as
\be\label{LAPSA}
\Delta \sta=\frac{2\sta+1}{\sta(\sta+1)}|\nabla \sta|^{2}.
\ee

\n The interesting thing about this expression is that it does not depend explicitly on $\alpha$. We note too that we have 
\be\label{STAGRAD}
<\frac{\nabla N}{N},\nabla \sta>=\frac{1}{2\alpha \sta(\sta+1)}|\nabla \sta|^{2}.
\ee

\n The crucial and obvious observation about the family $\{\sta\}$ is that given any open set $\Omega$ of compact closure $\bar{\Omega}\subset \Sigma\setminus H$ then $\sta$ converges uniformly in $C^{2}$ to $\st$ over $\bar{\Omega}$ as $\alpha\rightarrow 1$. Thus it follows from the limits of $\st$ at $H$ and infinity observed at the beginning that if $\max\{|\nabla \st|(q),q\in \Sigma\}>1$ then there is an $\epsilon>0$ such that for every $\alpha$ with $|\alpha -1|<\epsilon$ the function $|\nabla \sta|$ posses at least one local maximum greater than one. For a given $\alpha$ we will denote by $p_{\alpha}$ a point at which a local maximum of $\sta$ greater than one takes place.  

We will use Weitzenb\"ock's formula
\be\label{WEF}
\frac{1}{2}\Delta |\nabla \sta|^{2}=|\nabla \nabla \sta|^{2}+<\nabla \Delta \sta,\nabla \sta>+2<\frac{\nabla N}{N},\nabla \sta>^{2},
\ee

\n and we will use it evaluated at $p_{\alpha}$. We note first that for every vector $w\in T_{p_{\alpha}}\Sigma$ we have $<\nabla_{w}\nabla \sta,\nabla \sta>=0$. Because of this we have $|\nabla\nabla\sta|^{2}=|\nabla\nabla \sta|_{T_{p_{\alpha}}\Sigma}^{2}=|\nabla\nabla \sta|_{\nabla\sta(p_{\alpha})^{\perp}}$ where $\nabla\sta(p_{\alpha})^{\perp}$ is the perpendicular subspace to $\nabla \sta$ in $T_{p_{\alpha}}\Sigma$. Thus we have 
\ben
|\nabla \nabla \sta|^{2}(p_{\alpha})\geq \frac{1}{2} tr_{\nabla \sta(p_{\alpha})^{\perp}}\nabla \nabla\sta=\frac{1}{2}\Delta \sta (p_{\alpha}).
\een

\n This expression will be used in the first term on the right hand side of equation (\ref{WEF}). For the second instead we note from equation (\ref{LAPSA}) that
\ben
\nabla \Delta \sta \bvl_{p_{\alpha}}=-(\frac{1}{\sta^{2}}+\frac{1}{(\sta+1)^{2}})|\nabla \sta|^{2}\bvl_{p_{\alpha}}. 
\een

\n For the third term on the right hand side of equation (\ref{WEF}) we will use equation (\ref{STAGRAD}). All together gives for equation (\ref{WEF}) the expression
\ben
0\geq \frac{1}{2}\Delta |\nabla \sta|^{2}\bvl_{p_{\alpha}}\geq  |\nabla \sta|^{2}(\frac{(2\sta+1)^{2}}{2(\sta^{2}(\sta+1)^{2}}-\frac{\sta^{2}+(\sta+1)^{2}}{\sta^{2}(\sta+1)^{2}}+\frac{2}{4\alpha^{2}}\frac{1}{\sta^{2}(\sta+1)^{2}})\bvl_{p_{\alpha}}.
\een

\n Further expanding the term in parenthesis we obtain
\ben
0\geq \frac{1}{2}\Delta |\nabla \sta|^{2}\bvl_{p_{\alpha}}\geq \frac{|\nabla \sta|^{2}}{2\sta^{2}(\sta+1)^{2}}(-1+\frac{1}{\alpha})\bvl_{p_{\alpha}}.
\een

\n Choosing $\alpha$ such that $1-\epsilon <\alpha<1$ we get a contradiction. This finishes the proof of the Proposition. \ep

Define now $\delta=\st-\hat{\st}$. We will study $\delta$, and it will be shown that it has asymptotically positive Laplacian (in a barer sense). 

\begin{Proposition} Let $(\Sigma,\tg,\ln N)$ be an asymptotically flat static solution with regular and connected horizon $H$. The Laplacian of $\delta$ has the following asymptotic expression
\ben
\Delta \delta\leq \frac{-\delta}{(s+2m)^{2}}+O(\frac{1}{s^{3}}),
\een

\n in the barer sense.
\end{Proposition}

Note that $\delta\geq 0$. However note too that because there are sequences $\{p_{i}\}$ for which $\delta(p_{i})\rightarrow 0$, it cannot be said that $\Delta \delta$ becomes negative outside a sufficiently big compact set. The asymptotic expression is however still valid.

\n {\it Proof:}

\vs
Recall first the expression for $\Delta \hst$ in equation (\ref{SBF}). We find first the asymptotic expression for $|\nabla \hst|^{2}$. But observing that $\hst = 2m(\frac{1}{1-N^{2}}-1)$ it is easily deduced from the asymptotic expression of $N$ that $\nabla \hst=\nabla r+O(1/r^{2})$. Thus $|\nabla \hst|^{2}=1+(1/r^{2})=1+O(1/s^{2})$. 

Now subtract to the expression (\ref{SL}) with $P=1$ and $|\nabla \st|^{2}=1$, the expression (\ref{SBF}). That gives
\ben
\Delta \delta\leq \frac{2}{\st+2m}(1+\frac{m}{\st+2m})-\frac{2}{\hst+2m}(1+\frac{m}{\hst+2m})+O(\frac{1}{\st^{3}})=
\een
\ben
=\frac{-2\delta}{(\st+2m)(\hst+2m)}+2m\frac{(\hst^{2}-\st^{2})}{(\st+2m)^{2}(\hst^{2}+2m)^{2}}+O(\frac{1}{\st^{3}}).
\een

\n Thus 
\ben
\Delta \delta\leq \frac{-\delta}{(\st+2m)^{2}}+O(\frac{1}{\st^{3}}).
\een

\n as claimed.\ep
 
\vs 
We prove now a crucial property of $\delta$, namely that  it is Lipschitz ``at large scales". To explain the concept we need to introduce some terminology. Let $\{(r,\theta,\varphi)\}$ a be a coordinate system as in 
Proposition \ref{PBS2}. Let $D$ be the annulus in $\field{R}^{3}$, $D=\{(r,\theta,\varphi),1\leq r\leq2\}$. For any $\lambda>0$ sufficiently small consider the map from $D$ into $\Sigma$ given by $\bar{x}\rightarrow \bar{x}/\lambda$. Denote by $\delta_{\lambda}$ the pull-back of $\delta$ to $D$, namely $\delta_{\lambda}(\bar{x})=\delta(\bar{x}/\lambda)$. 
Let $\bar{x}_{1}$ and $\bar{x}_{2}$ be two points in $D$. Denote by $\phi(\bar{x}_{s},\bar{x}_{2})$ the angle formed by $\bar{x}_{1}$ and $\bar{x}_{2}$, namely $<\bar{x}_{1},\bar{x}_{2}>=|\bar{x}_{1}||\bar{x}_{2}|\cos \phi(\bar{x}_{1},\bar{x}_{2})$. We would like to show that there is $\lambda_{0}>0$ and $K>0$ such that $\delta_{\lambda}$ is Lipschitz with constant $K$ for any $0<\lambda<\lambda_{0}$. The next Proposition explains this property and two further that will also be needed later. It is perhaps the most technical, but otherwise straightforward Proposition of the article.
\begin{Proposition} Let $\delta=\st-\hst$. Then
\begin{enumerate} 
\item There exists $K>0$ and $\lambda_{0}>0$ such that for any $\bar{x}_{1},\bar{x}_{2}$ in $D$ and $0<\lambda<\lambda_{0}$ we have
\ben
|\delta_{\lambda}(\bar{x}_{1})-\delta_{\lambda}(\bar{x}_{2})|\leq K |\bar{x}_{1}-\bar{x}_{2}|,
\een
 
\item Let $\bar{x}_{1}$ and $\bar{x}_{2}$ be two points in $D$ belonging to the same radial line, namely $\bar{x}_{1}=\beta \bar{x}_{2}$. Then for any sequence $\{\lambda_{i}\}\downarrow 0$ we have
$|\delta_{\lambda_{1}}(\bar{x}_{1})-\delta_{\lambda_{i}}(\bar{x}_{2})|\rightarrow 0$.
\end{enumerate}
\end{Proposition}

\n {\it Proof:}

\vs
In $\Sigma$ consider a coordinate sphere $S_{r_{0}}=\{\bar{x}/ r(\bar{x})=r_{0}\}$ (where $\{\bar{x}\}$ is a coordinate system as in Proposition \ref{PBS2}). The distance function from $S_{r_{0}}$ to $H$ is Lipschitz, say with constant $K_{1}$, namely for any $q_{0}$, $q_{1}$ in $S_{r_{0}}$ we have
$|\st(q_{0})-\st(q_{1})|\leq K_{1} |\phi(q_{0},q_{1})|$. 

Let now $\bar{x}_{1}$ be a point in $D$. Let $\lambda$ such that $|\bar{x}_{1}|/\lambda >>r_{0}$. Denote $p_{1}=\bar{x}_{1}/\lambda$. Let $\gamma_{1}$ be the length minimizing geodesic joining $\bar{x}_{1}/\lambda$ to $H$. Let $q_{1}$ be the point of intersection of $\gamma_{1}$ with $S_{r_{0}}$. Consider a rotation of angle $\phi_{0}$ in $\field{R}^{3}$, denote it by $R_{\phi_{0}}$. Also denote by $p_{2}=R_{\phi_{0}}(p_{1})$, $\gamma_{2}=R_{\phi_{0}}(\gamma_{1})$ and $q_{2}=R_{\phi_{0}}(q_{1})$. Let $l_{1}$ be the length of $\gamma_{1}$ between $p_{1}$ and $q_{1}$ and let $l_{2}$ be the length between $p_{2}$ and $q_{2}$ of $\gamma_{2}$. 

We will show first that there is a constant $K_{2}>0$ independent on $\lambda$ such that $|l_{1}-l_{2}|\leq K_{2} |\phi_{0}|$. Note that in the coordinate system $\{\bar{x}\}$ we have $\tg=\tg_{S}+O(1/r^{3})$. Suppose $\gamma_{1}$ is parameterized with respect to the arc-length, $\bar{\st}$, provided by the Schwarzschild metric $\tg_{S}$. Let $l(\phi)=l(R_{\phi}(\gamma_{1}))$, where $0<\phi<\phi_{0}$. Then we have  
\be\label{DERL}
|\partial_{\phi}l|=|\int_{\bar{\st}_{0}=0}^{\bar{\st}_{1}}\frac{\tg(\nabla_{\partial_{\phi}}\gamma',\gamma')}{\tg(\gamma',\gamma')^{\frac{1}{2}}}d\bar{\st}|. 
\ee 

\n Moreover 
\ben
\tg(\nabla_{\partial_{\phi}}\gamma',\gamma')=\tg((\nabla_{\partial_{\phi}}-\nabla^{S}_{\partial_{\phi}})\gamma',\gamma')+(\tg-\tg_{S})(\nabla^{S}_{\partial_{\phi}}\gamma',\gamma')+\tg_{S}(\nabla^{S}_{\partial_{\phi}}\gamma',\gamma').
\een

\n We note now that the last term on the right hand side of the previous equation is zero, and the first two terms on the right hand side are $O(1/\bar{\st}^{2})$. Using this in equation (\ref{DERL}) we get that $|l_{1}-l_{2}|\leq K_{2}|\phi_{0}|$ as desired.  

We have now
\ben
\st(p_{2})\leq l_{2}+\st(q_{2})\leq l_{1}+\st(q_{1})+K_{1}|\phi_{0}|+K_{2}|\phi_{0}|=\st(p_{1})+(K_{1}+K_{2})|\phi_{0}|.
\een

\n Because $p_{1}$ and $\phi$ are arbitrary we have
\ben
|\st(p_{1})-\st(p_{2})|\leq K|\phi_{0}|.
\een

\n Thus for any $\bar{x}_{1}$ and $\bar{x}_{2}$ in $D$ of equal norm, $|\bar{x}_{1}|=|\bar{x}_{2}|$, and $\lambda$ (sufficiently small), we have 
\be\label{IIU}
|\delta_{\lambda}(\bar{x}_{1})-\delta_{\lambda}(\bar{x}_{2})|=|\st(\frac{\bar{x}_{1}}{\lambda})-\frac{|\bar{x}_{1}|}{\lambda}+2m-\st(\frac{\bar{x}_{2}}{\lambda})+\frac{|\bar{x}_{2}|}{\lambda}-2m|=|\st(\frac{\bar{x}_{1}}{\lambda})-\st(\frac{\bar{x}_{2}}{\lambda})|\leq K|\phi(\bar{x}_{1},\bar{x}_{2})|.
\ee

We continue with an observation. Recall that the Ricci curvature of $\tg$ decays, in $r$, as $O(1/r^{3})$ (in facts it decays as $1/r^{4}$). Consider the annulus $D_{\lambda}=\{\bar{x},\ \lambda^{1/12}\leq |\bar{x}|\leq 2\}$ and consider the map from $D_{\lambda}$ into $\Sigma$ given by $\bar{x}\rightarrow \bar{x}/\lambda$. Let $\tg_{\lambda}$ be the pull-back of the metric $\tg$ under this map. The from the fact that $|{\tt Ric}|$ decays as $O(1/r^{3})$ we get $\sup\{|{\tt Ric}_{\tg_{\lambda}}(\bar{x})|_{\tg_{\lambda}}/ \bar{x}\in D_{\lambda}\}=O(\lambda^{\frac{1}{4}})$. From this it follows that, as $\lambda$ tends to zero, and therefore as $D_{\lambda}$ tends to the closed ball of radius two minus the origin, the metrics $\tg_{\lambda}$ converge in $C^{1,\beta}$ (for any $0<\beta<1$) to the flat metric over any fixed annulus $D_{\lambda_{1}}$, $0<\lambda_{1}<2$. Thus for any $\bar{x}\in D$ and sequence $\{\lambda_{i}\}\downarrow 0$, length minimizing geodesics, $\gamma_{p}$, joining $p=\bar{x}/\lambda$ to $H$ converge in $C^{1}$ over any $D_{\lambda_{1}}$ to the radial line  passing through $\bar{x}$. 

What we would like to know now is the ``rate" at which the geodesics approach the radial lines. More precisely, we will study the $\tg_{S}$-angle $\xi$, formed by $\partial_{r}$ and $\gamma'$ at any point along $\gamma$. To this respect we proceed as follows. Consider the rotational killing fields $X$ of the Schwarzschild solution. For every $X$, we have $|X|_{\tg}=r(1+O(1/r))$. Given one of the $X$'s, we compute, along the geodesic $\gamma_{p}$ (again $p=\bar{x}/\lambda$)
\ben
\tg(\gamma',X)'=\tg(\gamma',\nabla_{\gamma'}X)=\tg(\gamma',(\nabla_{\gamma'}-\nabla^{S}_{\gamma'}) X)+\tg_{S}(\gamma',\nabla^{S}_{\gamma'}X)+(\tg-\tg_{S})(\gamma',\nabla^{S}_{\gamma'} X).
\een     

\n The second term on the right hand side of the previous equation is zero, while the other two are of the order $O(1/r^{2})=O(1/\st^{2})$. Let $q$ be the first point where $\gamma_{p}$ reaches the radial sphere $S_{r_{0}}$ ($r_{0}$ is fixed) and let $p_{1}$ be any intermediate point between $p$ and $q$. Integrate now $\tg(\gamma',X)'$ (with respect to the $\tg$ arc-length, $\st$) between $\st(p_{1})$ and the value of $\st(q)$ using the estimate we have found before for $\tg(\gamma',X)$
to get
\ben
|\tg(\gamma',X)(p_{1})-\tg(\gamma',X)(q)|\leq c_{1},  
\een

\n where $c_{1}$ is a constant independent on $p_{1}$ and $q$. Note that this inequality is valid for any rotational Killing field $X$. Observing that rotational killing fields at $S_{r_{0}}$ have bounded norm, we get
\ben
|\tg(\gamma',X)(p_{1})|\leq c_{2},
\een

\n where $c_{2}$ is a constant. Moreover
\ben
\tg_{S}(\gamma',X)=\tg(\gamma',X)+(\tg_{S}-\tg)(\gamma',X)=\tg(\gamma',X)+O(1/r).
\een

\n Thus we have
\ben
|\tg_{S}(\gamma',X)|\leq c_{3},
\een

\n where $c_{3}$ is a constant. Pick now the rotational killing field $X$ which is collinear, at
$p_{1}$, to the component of $\gamma'$, $\tg_{S}$-perpendicular to $\partial_{r}$. Let $\xi$ be the $\tg_{S}$-angle formed by $\partial_{r}$ and $\gamma'$. We have 
\ben
|\tg_{S}(\gamma',X)(p_{1})|=|X|_{\tg_{S}}(p_{1})||\gamma'|_{\tg_{S}}|\sin \xi(p_{1})|\leq c_{4},
\een

\n where $c_{4}$ is a constant. So we get
\ben
|\sin \xi|\leq \frac{c_{5}}{r},
\een

\n where $c_{5}$ is a constant. We have
\be\label{RSI}
\frac{dr}{d\st}=\tg_{S}(\nabla^{S} r,\gamma')=1+O(1/r^{2})=1+O(1/\st^{2}).
\ee

\n We will use this inequality in what follows. Let $\bar{x}_{1}$ be a point in $D$. Let $p_{1}=\bar{x}_{1}/\lambda$ and let $\gamma$ be a geodesic minimizing the length between $p_{1}$ and $H$. Let $p_{2}$ be a point in $\gamma$ such that $p_{2}=\bar{x}_{2}/\lambda$ with $\bar{x}_{2}$ in $D$. Integrating (\ref{RSI}) between $\st(p_{1})$ and $\st(p_{2})$ we get
\ben
r(p_{1})-\st(p_{1})=r(p_{2})-\st(p_{2})+|\bar{x}_{1}-\bar{x}_{2}|O(\lambda).
\een

\n Therefore
\be\label{IID}
|\delta_{\lambda}(\bar{x}_{1})-\delta_{\lambda}(\bar{x}_{2})|=|\bar{x}_{1}-\bar{x}_{2}|O(\lambda).
\ee

We are ready to prove the Proposition. Let $\bar{x}_{1}$ and $\bar{x}_{2}$ be two points in $D$. Let $p_{1}=\bar{x}_{1}/\lambda$ and $p_{2}=\bar{x}_{2}/\lambda$. Let $p_{3}=\bar{x}_{3}/\lambda$ be the point of intersection of the length minimizing geodesic joining $p_{1}$ to $H$ and the coordinate sphere $S_{|\bar{x}_{2}/\lambda|}$. From (\ref{IIU}) and (\ref{IID}) we get
\ben
|\delta_{\lambda}(\bar{x}_{1})-\delta_{\lambda}(\bar{x}_{2}|\leq |\delta_{\lambda}(\bar{x}_{1})-\delta_{\lambda}(\bar{x}_{3})|+|\delta_{\lambda}(\bar{x}_{3}-\delta_{\lambda}(x_{2})|\leq |\bar{x}_{1}-\bar{x}_{3}|O(\lambda)+K\phi(\bar{x}_{3},\bar{x}_{2}).
\een

\n As $|\bar{x}_{1}-\bar{x}_{3}|\leq c_{6} d_{D}(\bar{x}_{1},\bar{x}_{3})$, for some constant $c_{6}$, the {\it item 1} of the Proposition follows. {\it Item 2} follows from the fact that $O(\lambda)\rightarrow 0$, as $\lambda\rightarrow 0$.\ep

\vs
The following direct implication will be crucial for the discussion that follows.

\begin{Corollary}
For any sequence $\{\lambda_{i}\}$ such that $\lambda_{i}\downarrow 0$, there exists a subsequence $\{\lambda_{i_{k}}\}\downarrow 0$ and a Lipschitz function $\delta_{0}$ (depending on $\{\lambda_{i_{k}}\}$) for which $\delta_{\lambda_{i_{k}}}$ converges uniformly to $\delta_{0}$ on $D$. The function $\delta_{0}$ is constant on radial lines.
\end{Corollary} 

We would like now to prove that the coordinate-distance lag $\bar{\delta}(\{p_{i}\})$ of any divergent sequence $\{p_{i}\}$ is zero. Naturally, this is the same as saying that $\delta$ converges uniformly to zero at infinity. If this is not the case, then it is simple to see, arguing by contradiction, that we would be in the following situation. There would exist $\{\lambda_{i}\}$ with $\lambda_{i}\downarrow 0$ such that $\delta_{\lambda_{i}}$ converges uniformly to a Liptschitz function function $\delta_{0}$ and there would exist points $x,y$ in $D$ for which $\delta_{0}(x)=0$, $|x|=3/2$ and $\delta_{0}(y)>0$, $|y|=3/2$ and $|x-y|<1/2$. Assume we are in such situation. Define in $D$ the Euclidean balls $B_{x}=B(x,|x-y|)$ and $B_{y}=B(y,\xi)$ where $\xi$ is small enough to have $\delta_{0}|_{B_{y}}>c_{1}>0$, where $c_{1}$ is a constant. Following ~\cite{MR2243772} (pg. 258) we can find a function $h$ on $\bar{B}_{x}$ such that
\begin{enumerate}
\item $h\bvl_{(\partial(B_{x})\setminus B_{y})}<c_{2}<0$, where $c_{2}$ is a constant,
\item $h(x)=0$,
\item $\Delta_{g_{\lambda_{i}}} h \bvl_{\bar{B}_{x}}> c_{3}>0$, where $c_{3}$ is a constant and $g_{\lambda_{i}}$ is the scaled metric $\lambda_{i}^{2}g$.
\end{enumerate}

\n Note that the scaled metrics $\lambda_{i}^{2}g$ converge (in $C^{\infty}$) to the flat Euclidean metric. 
As $\delta_{\lambda_{i}}$ converges uniformly to $\delta_{0}$ we deduce that there is $\mu_{0}>0$ such that for any $0<\mu\leq \mu_{0}$ (and $i\geq i_{0}(\mu_{0})$) we have $(-\delta_{\lambda_{i}}+\mu h )|_{\partial B_{x}}<\mu c_{4}<0$, where $c_{4}$ is a constant. We also have $\lim (-\delta_{\lambda_{i}}(x)+\mu h(x))\rightarrow 0$. It follows that having chosen $i_{1}$ big enough, the function $-\delta_{\lambda_{i}}+\mu h$, ($\mu\leq \mu_{0}$), for $i\geq i_{1}$ has a maximum on $B_{x}$. Denote it by $z_{i}$. If the function $\st$ were to be smooth at $z_{i}/\lambda_{i}$ and therefore $-\delta_{\lambda_{i}}+\mu h$ were smooth at $z_{i}$ then one would get a contradiction to the maximum principle, as for $i$ sufficiently big, one would have
\ben
\Delta_{g_{\lambda_{i}}}(-\tilde{\delta}_{\lambda_{i}}+\mu h )(z_{i})\geq  \frac{\mu c_{3}}{2}>0.
\een

We explain now how to use Proposition \ref{LSH} to overcome the case when $z_{i}$ are not smooth points of $\st$. One can replace $\st$ by $\st_{\Gamma_{i}}$, for a suitable $\{\Gamma_{i}\}\downarrow 0$, in the expression $\delta_{\lambda_{i}}(x)=(\st-\hst)(x/\lambda_{i})$ in such a way that the new expression $(-(\st_{\Gamma_{i}}-\hst)+\mu h)(x/\lambda_{i}$), has a maximum $\tilde{z}_{i}$ on $B_{x}$. Further, by Proposition \ref{LSH} one can replace $\st_{\Gamma_{i}}$ by $\tilde{\st}_{\Gamma_{i}}$ in such a way that the new expression $\tilde{\delta}_{\lambda_{i}}(x)=(\tilde{\st}_{\Gamma_{i}}-\hst)(x/\lambda_{i})$  satisfies 
\begin{enumerate}
\item $-\tilde{\delta}_{\lambda_{i}}(x)=-(\tilde{\st}_{\Gamma_{i}}-\hst)(x/\lambda_{i})\leq -(\st_{\Gamma_{i}}-\hst)(x/\lambda_{i}$),
\item $-\tilde{\delta}_{\lambda_{i}}(\tilde{z}_{i})=(\st_{\Gamma_{i}}-\hst)(\tilde{z}_{i}/\lambda_{i}$), and thus $-\tilde{\delta}_{\lambda_{i}}+\mu h$ has a maximum at $\tilde{z}_{i}$ on $B_{x}$.
\item $\Delta_{g_{\lambda_{i}}}( -\tilde{\delta}_{\lambda_{i}}+\mu h )(\tilde{z}_{i})\geq  \frac{\mu c_{3}}{2}.$
\end{enumerate}

\n These three facts now contradict the maximum principle.\ep

We have thus proved

\begin{Proposition}\label{DELZ} Let $(\Sigma,\tg,\ln N)$ be an asymptotically flat static solution with regular and connected horizon. Then for any divergent sequence $\{p_{i}\}$, the coordinate-distance lag $\bar{\delta}(\{p_{i}\})$ is zero. 
\end{Proposition}
   
\subsubsection{Area and volume comparison.}\label{AVC}   
   
\begin{Proposition}\label{VOLZ} Let $(\Sigma,\tg,\ln N)$ be an asymptotically flat static solution with regular an connected horizon. Consider a sequence $\{\Gamma_{i}\}\downarrow 0$. Let ${\mathcal{F}}_{\Gamma_{i}}$ be the congruence of length minimizing geodesics to $H_{\Gamma_{i}}$. Then for every $L>0$ we have
\ben
Vol(\cup_{\gamma\in {\mathcal{F}}_{\Gamma_{i}}, l(\gamma)\leq L}\{\gamma\})\rightarrow 0,
\een

\n as $\Gamma_{i}\downarrow 0$. Above $\{\gamma\}$ means the set of points in $\gamma$.
\end{Proposition}

\n {\it Proof:}

\vs
The first goal to achieve is to make the monotonicity of ${\mathcal{M}}$ to look like a {\it comparison of areas} and consequently a {\it comparison of volumes}. Let $\{\Gamma_{i}\}\downarrow 0$. Consider for each $\Gamma_{i}$ the congruence ${\mathcal{F}}_{\Gamma_{i}}$ of length minimizing geodesics to $H_{\Gamma_{i}}$. We will work outside the locus at all times. Let $dA$ be the element of area of the level sets of the congruence. Let $\st_{\Gamma_{i}}$ be the distance function to $H_{\Gamma_{i}}$. Then 
\ben
\theta=\frac{1}{A}\frac{dA}{d\st_{\Gamma_{i}}}.
\een

\n Let $\gamma$ be a geodesic in ${\mathcal{F}}_{\Gamma_{i}}$. Consider ${\mathcal{M}}_{a}$ with $a=2m$ over $\gamma$. Denote by ${\mathcal{M}}_{\Gamma_{i}}$ the value of ${\mathcal{M}}$ at the initial point of $\gamma$ in $H_{\Gamma_{i}}$. Then from the monotonicity of ${\mathcal{M}}$ we have
\ben
(\frac{1}{2A}\frac{dA}{d\st_{\Gamma_{i}}}(\st_{\Gamma_{i}}+2m)^{2}-(\st_{\Gamma_{i}}+2m))N^{2}\leq {\mathcal{M}}_{\Gamma_{i}}.
\een    

\n Rearranging terms we get
\ben
\frac{d}{d\st_{\Gamma_{i}}}(\frac{dA}{(\st_{\Gamma_{i}}+2m)^{2})})\leq \frac{2{\mathcal{M}}_{\Gamma_{i}}}{N^{2}(\st_{\Gamma_{i}}+2m)^{2}}dA.
\een

\n We thus get
\ben
\frac{d}{d\st_{\Gamma_{i}}} \ln \frac{dA}{(\st_{\Gamma_{i}}+2m)^{2}}\leq \frac{2{\mathcal{M}}_{\Gamma_{i}}}{N^{2}(\st_{\Gamma_{i}}+2m)^{2}}.
\een

\n Integrating we obtain
\be\label{ACOM}
\frac{dA}{(\st_{\Gamma_{i}} +2m)^{2}}\leq \frac{dA_{0}}{(2m)^{2}}\exp (\int_{0}^{\st_{\Gamma_{i}}}\frac{2{\mathcal{M}}_{\Gamma_{i}}}{N^{2}(\st_{\Gamma_{i}}+2m)^{2}}d\st_{\Gamma_{i}}).
\ee

\n where $dA_{0}$ is the element of area of $H_{\Gamma_{i}}$. Recalling that $N^{2}=\hst/(\hst+2m)$ it is clear that we need an estimation of $\hst$ in terms of $\st_{\Gamma_{i}}$ to have an inequality in terms of $\st_{\Gamma_{i}}$ only. We advocate to that in the following lines. We explain first how to get a relation between $\st$ and $\st_{\Gamma_{i}}$ and then we explain how to obtain one in terms of $\hat{\st}$ and $\st_{\Gamma_{i}}$.

First recall from (\ref{C4}) that for any point $q$ in $H_{\Gamma_{i}}$ we have (for $\Gamma_{i}$ small enough) that $\st(q)=\hst(q)+O(\hst^{\frac{3}{2}})$. Now let $p$ be a point in $\gamma$. Then we have
$\st(p)\leq \st_{\Gamma_{i}}(p)+\st(q)$, where here $q$ is the initial point of $\gamma$ at $H_{\Gamma_{i}}$. Thus $\st(p)\leq \st_{\Gamma}(p)+(1+\epsilon)\hst(p)$ where $\epsilon=O(\hst(p)^{\frac{1}{2}})$. On the other hand let $\bar{\gamma}$ be a length minimizing geodesic joining $p$ to $H$. Let $\bar{q}$ be the point of intersection to $H_{\Gamma_{i}}$. Then we have 
\ben
\st(p)=dist(p,\bar{q})+\st(\bar{q})\geq \st_{\Gamma_{i}}(p)+\hst(\bar{q})+O(\hst({\bar{q}})^{\frac{3}{2}})\geq\st_{\Gamma_{i}}(p)+(1-\epsilon)\hst(q),
\een

\n where $\epsilon=O(\hst(q)^{\frac{1}{2}})$. Thus for every point $p$ in $\gamma$ we have
\ben
(1-\epsilon)\hst_{0}+\st_{\Gamma_{i}}(p)\leq \st(p)\leq \st_{\Gamma_{i}}(p)+(1+\epsilon)\hst_{0},
\een

\n where we have made $\hst_{0}=\hst(q)$ to simplify the notation. This establishes the relation between $\st$ and $\st_{\Gamma_{i}}$. We obtain now the desired relation between $\st_{\Gamma_{i}}$ and $\hst$. We will keep the notation as before. Precisely, $\gamma$ will be length minimizing geodesic segment to $H_{\Gamma_{i}}$ and $q$ and $q_{1}$ will be its initial and final points. From Proposition \ref{GRADUN}, we know that $|\nabla \hst|\leq 1$ therefore for any point $p$ between $q$ and $q_{1}$ we have
\begin{align*}
&\hst(q_{1})-\hst(p)\leq \st_{\Gamma_{i}}(q_{1})-\st_{\Gamma_{i}}(p),\\
&\hst(p)-\hst(q)\leq \st_{\Gamma_{i}}(p).
\end{align*}
   
\n Using this we have
\ben
(1+\epsilon)\hst_{0}\geq \hst(q)\geq \hst(p)-\st_{\Gamma_{i}}(p)\geq \hst(q_{1})-\st_{\Gamma_{i}}(q_{1})\geq
\hst (q_{1})-\st(q_{1})+(1-\epsilon)\hst_{0}.
\een

\n Now if $\st(q_{1})\geq \bar{L}$ and $\bar{L}=\bar{L}(\Gamma_{i})$ is big enough we have $\hst(q_{1})-\st(q_{1})\geq -\epsilon \hst_{0}$. As a result we have the relation
\be\label{SSHR}
(1+\epsilon)\hst_{0}\geq \hst(p)-\st_{\Gamma}(p)\geq (1-2\epsilon)\hst_{0}.
\ee

We have now all the elements to proceed with the proof of the Proposition. Consider the set of the initial points on $H_{\Gamma_{i}}$ of the geodesics in ${\mathcal{F}}_{\Gamma_{i}}$ whose lengths are greater than $\bar{L}(\Gamma_{i})$. Denote such set by $\Omega_{\Gamma_{i}}$. We will show now that as $\Gamma_{i}\downarrow 0$, and therefore as $H_{\Gamma_{i}}$ approaches $H$, the area of $\Omega_{\Gamma_{i}}$ with respect to the area element induced from $g$ tends to the total area of the horizon $H$.     

Consider the argument in the exponential function of (\ref{ACOM}) with the upper limit of integration equal to $\bar{L}$. Using the relation (\ref{SSHR}) we obtain
\begin{align*}
\int_{0}^{\bar{L}}\frac{{\mathcal{M}}_{0}}{N^{2}(\st_{\Gamma_{i}}+2m)^{2}}d\st_{\Gamma_{i}}&=\int_{0}^{\bar{L}} \frac{{\mathcal{M}}_{0}(\hst+2m)}{\hst^{2}(\st_{\Gamma_{i}}+2m)^{2}}d\st_{\Gamma_{i}}\\
&\leq\int_{0}^{\bar{L}}\frac{{\mathcal{M}}_{0}(\st_{\Gamma_{i}}+2m+(1+\epsilon)\hst_{0})}{(\st_{\Gamma_{i}}+(1-2\epsilon)\hst_{0})(\st_{\Gamma_{i}}+2m)^{2}}d\st_{\Gamma_{i}}. 
\end{align*}

\n This last integral can be further split into
\ben
\int_{0}^{\bar{L}} \frac{{\mathcal{M}}_{0}}{(\st_{\Gamma_{i}}+(1-2\epsilon)\hst_{0})(\st_{\Gamma_{i}}+2m)}d\st_{\Gamma_{i}}+R(\hst_{0}),
\een

\n where $R(\hst_{0})$ is an expression which is easily seen to tend to zero as $\hst_{0}$ tends to zero. 
We integrate now equation (\ref{ACOM}) in $dA$. After integrating in $dA$, the left hand side tends to $4\pi$ for a suitable divergent sequence of $\bar{L}$'s. The right hand side is easily integrated to be (discard the term $R(\hst_{0})$)
\ben
\int_{\Omega_{\Gamma_{i}}} \frac{\hst_{0}}{(\hst_{0}+2m)(2m)^{2}}(\frac{2m}{(1-2\epsilon)\hst_{0}})^{\frac{2{\mathcal{M}}_{0}}{2m-(1-2\epsilon)\hst_{0}}}dA_{g},
\een

\n where $dA_{g}=N^{2}dA_{0}=\frac{\hst_{0}}{\hst_{0}+2m}dA$ is the element of area induced on $H_{\Gamma_{i}}$ from the metric $g$. As a result we get the inequality
\be\label{INEF}
4\pi \leq \frac{\limsup A(\Omega_{\Gamma_{i}})}{4m^{2}} \limsup \hst_{0}^{\frac{ 2{\mathcal{M}}_{0}-2m+(1-2\epsilon)\hst_{0}}{2m-(1-2\epsilon)\hst_{0}}}.
\ee

\n Now, from the proof of Proposition \ref{VMH} it is seen that $|{\mathcal{M}}_{0}-m|\leq c_{1}\hst^{\frac{1}{2}}_{0}$ where $c_{1}$ is a positive constant.  Thus we get
\ben
\hst_{0}^{\frac{ 2{\mathcal{M}}_{0}-2m+(1-2\epsilon)\hst_{0}}{2m-(1-2\epsilon)\hst_{0}}}\leq \hst_{0}^{c_{2}\hst_{0}^{\frac{1}{2}}}\rightarrow 1,\ {\rm as}\ \hst_{0}\rightarrow 0,
\een

\n where $c_{2}$ is a positive constant. Therefore we get from this and equation (\ref{INEF})
\ben
16\pi m^{2}\leq \limsup A(\Omega_{\Gamma_{i}})\leq A=16\pi m^{2},
\een

\n where $A$ is the area of the horizon. Thus $\lim\sup A(\Omega_{\Gamma_{i}})=A$. This was the crucial estimate. From it, it will follow that for any $L<\infty$ fixed, there is a subsequence $\Gamma_{i_{j}}$ such that the area of the set of initial points in $H_{\Gamma_{i_{j}}}$ of the geodesics in ${\mathcal{F}}_{\Gamma_{i_{j}}}$ whose length is less or equal than $L$, tends actually to zero. This would finish the proof of the Proposition. We do that now. For every $j$, denote by $\Omega_{L,\Gamma_{i_{j}}}$ such set. For every $q$ in $\Omega_{L,\Gamma_{i_{j}}}$ let $\gamma_{q}$ be the corresponding geodesic in ${\mathcal{F}}_{\Gamma_{i_{j}}}$ whose total length is less than or equal to $L$. Denote by $U_{L,\Gamma_{i_{j}}}$ the union $U=\cup_{q\in\Omega_{L,\Gamma_{i_{j}}}} \{\gamma_{q}\}$.  Now, recalling that $dV'=dA$, integrating equation (\ref{ACOM}), and following the same treatment at the horizon as before gives
\ben
Vol_{\tg}(U_{L,\Gamma_{i_{j}}})\leq c(L)A_{g}(\Omega_{L,\Gamma_{i_{j}}}).
\een

\n Note that in this equation, the volume is found with $\tg$ while the area is found with $g$. As $A(\Omega_{i_{j}})\rightarrow 0$, the Proposition follows.\ep    

\vs
The Proposition before has the following quite important Corollary.
\begin{Corollary}\label{CORFIN} Let $(\Sigma,\tg,\ln N)$ be an asymptotically flat static solution with regular an connected horizon. Then
\begin{enumerate}
\item $\st=\hst$ and therefore $\st$ is smooth.
\item $|\nabla \hst|^{2}=1$.
\item The integral curves of $\nabla \hst$ are geodesics minimizing the length between any two of its points.
\item The set of integral curves of $\nabla \hst$ form an integrable congruence of geodesics.
\end{enumerate}
\end{Corollary}

\n {\it Proof:}

\vs
	Let $p \in \Sigma\setminus H$. Let $\{\Gamma_{i}\}$ such that $\Gamma_{i}\downarrow 0$. Following Proposition \ref{VOLZ} there is a sequence $\{\gamma_{i}\}$ of length minimizing geodesics to $H_{\Gamma_{i}}$ with initial point $q_{i}$ (at $H_{\Gamma_{i}}$), $l(\gamma_{i})\rightarrow \infty$ and $\gamma_{i}(s(p))\rightarrow p$. Let $p_{i}$ be either the end point of $\gamma_{i}$ or, if $l(\gamma_{i})=\infty$, a point on $\gamma_{i}$ such that $s(p_{i})\rightarrow \infty$. We have 
\be\label{SSTD}
\hst(p_{i})-\hst(q_{i})=\int_{\bar{s}(q_{i})=0}^{\bar{s}(p_{i})}<\nabla\hst,\gamma'>d\bar{s}=\bar{s}(p_{i})-\bar{s}(q_{i})-\int_{\bar{s}(q_{i})}^{\bar{s}(p_{i})}(1-<\nabla \hst,\gamma'>)d\bar{s}.
\ee

\n where $\bar{\st}$ is the arc-length. But by Proposition \ref{DELZ} we have $\lim \delta(p_{i})=\st(p_{i})-\hst(p_{i})=0$ and thus we have $\lim \bar{\st}(p_{i})-\hst(p_{i})=0$ (note that $\lim |\st(p_{i})-\bar{\st}(p_{i})|=0$). By Proposition \ref{GRADUN} we have
$(1-<\nabla \hst,\gamma'>)\geq 0$, thus from equation (\ref{SSTD}) we get
\ben
0\leq \lim \int (1-<\nabla \hst,\gamma'>)d\bar{\st}=0,
\een

\n This shows $|\nabla \st|(p)=1$. Moreover we have 
\ben 
\hst(p)=\lim \hst(p_{i})-\hst(q_{i})=\lim\ \bar{\st}(p_{i})-\bar{\st}(q_{i})-\int_{\bar{\st}(q_{i})}^{\bar{\st}(p_{i})} (1-<\nabla \hst,\gamma'>)d\bar{\st}=\lim \bar{\st}(p_{i})=\st(p).
\een

\n Because $p$ is an arbitrary point we have thus proved {\it items 1,2} of the Proposition.

To prove the third {\it item} we proceed like this. Let $\gamma$ be an integral curve of $\nabla \hst$ with initial point $p$ and final point $q$. Suppose that $\gamma$ does not minimize the distance between $p$ and $q$, namely that there is another curve $\tilde{\gamma}$ joining $p$ and $q$ and having smaller length. Then
\ben
\st(q)=\st(p)+(\st(q)-\st(p))=\st(p)+l(\gamma)<\st(p)+l(\tilde{\gamma})\leq \st(q).
\een

\n which is a contradiction. 

{\it Item 4} of the Proposition follows directly from the fact that the congruence is orthogonal to the level set of any regular value of $\st$. \ep 

\subsubsection{The uniqueness of the Schwarzschild solutions.}

\begin{Theorem}\label{TUSFIN} Let $(\Sigma,\tg,\ln N)$ be an asymptotically flat static solution with regular an connected horizon. Then the solutions is a Schwarzschild solution of positive mass. 
\end{Theorem}

\n {\it Proof:}

\vs
By Corollary \ref{CORFIN} the set of integral curves of $\nabla \hst$ is an integrable congruence of geodesics. Recalling that $|\nabla \hst|=1$ and $\Delta\hst=\theta$, where $\theta$ is the mean curvature of the congruence. Using these facts in equation (\ref{SBF} we get that
\ben
{\mathcal{M}}_{a=2m}=(\frac{\theta(\st+2m)^{2}}{2}-(\st+2m))N^{2}=m,
\een

\n over any geodesic of the congruence. The conclusion that the solution is the Schwarzschild solution follows from Proposition \ref{COM} and the Remark after it.\ep 

\bibliographystyle{plain}
\bibliography{Master}

\end{document}